\documentstyle[axodraw]{cjp3}

\begin{document}

\title{\bf Chiral Symmetry and Low Energy Pion-Nucleon Scattering}

\authori{Sidney A.\ Coon\thanks{Electronic-mail address: coon@nmsu.edu}}

\addressi{Physics Department, New Mexico State University, Las Cruces, NM 
 88003}

\authorii{}       
\addressii{}
\authoriii{}      
\addressiii{}     

\headtitle{Chiral Symmetry $\ldots$}
\headauthor{Sidney A. Coon}

\evidence{A}
\daterec{XXX}    
\cislo{0}  \year{1998}
\setcounter{page}{1}

\maketitle
\begin{abstract} 
In these lectures, I examine the effect of the meson factory $\pi$N
data  on the current algebra/PCAC program which describes  chiral
symmetry breaking in this system.  After historical remarks on the
current algebra/PCAC  versus chiral Lagrangians approaches to chiral
symmetry, and description of the need for $\pi$N amplitudes with
virtual (off-mass-shell) pions in nuclear force models and other
nuclear physics problems, I begin with kinematics and isospin aspects
of the invariant amplitudes.  A detailed introduction to the hadronic
vector and axial-vector currents and the hypothesis of partially
conserved axial-vector currents (PCAC) follows. I review and test
against contemporary data the PCAC predictions of the 
Goldberger-Treiman relation, and the Adler consistency condition for a
$\pi$N amplitude.  Then comes a detailed description of the current
algebra Ward-Takahashi identities in the chiral limit and a brief
account of the on-shell current algebra Ward-Takahashi identities.  The
latter identities form the basis of so-called current algebra models of
$\pi$N scattering.  I then test these models  against the
contemporary empirical $\pi$N amplitudes  extrapolated into the
subthreshold region via dispersion relations.  The scale and the $t$
dependence of the ``sigma term" is determined by the recent data.

\end{abstract}

\section{Introduction}
	The implementation of chiral symmetry  in hadronic physics
began around 1960. Its consequences were examined with  two basic
approaches. One is based on the concept of a partially conserved
axial-vector current (PCAC) coupled with the algebra of vector and
axial-vector hadronic currents.  This current algebra is expressed as 
equal time commutation relations (the analogue of angular momentum
commutation relations in quantum mechanics). The other  (Lagrangian
form) is based on chiral Lagrangians with small explicit chiral
symmetry breaking terms.  A famous example of the latter is the linear
sigma model of Gell-Mann and Levy \cite{GML60} which explicitly
exhibits {\em both} PCAC and the current algebra. To adapt J. B. S.
Haldane's famous remark about the Deity and beetles, chiral symmetry
seems to be inordinately fond of pions.  Single pion exchange accounts
for about 70\% of the binding of light nuclei~\cite{binding} (and
perhaps all nuclei) and pions make up  the most prominent non-nucleon
degree of freedom in nuclear physics.   Thus it should not be
surprising that chiral symmetry was  applied to nuclear physics  as
soon as 1967. It is the aim of these lectures, motivated by the nuclear
physics problems briefly mentioned below, to discuss the PCAC-current
algebra approach  with the aid of contemporary experimental knowledge
of the low energy pion nucleon interaction.  Relationships between the
two approaches to chiral symmetry breaking will be mentioned when
useful. This introductory lecture is primarily for motivation and  will
freely employ  undefined concepts that will be defined and derived in
detail in the following lectures.

	 One of the earliest and boldest uses, in nuclear physics, of
the PCAC form of chiral symmetry was the relationship obtained, by
Blin-Stoyle and Tint, between the $\beta$-decay pion-exchange operator
and a phenomenological two-body (nucleons) pion production
operator~\cite{BST67}.  With this relation, they attempted to analyze
the process $p + p \rightarrow d + \pi^+$ using  two-body
terms obtained from a comparison of $\beta$-decay  of the tritium
nucleus and $\beta$-decay of the neutron.  Neither the pion production
data, the three-body wavefunction, nor the $ft$ values of the two
$\beta$-decays were known in the 60's well enough to obtain a
quantitative conclusion.  Nearly the same technique was used 30 years
later  to obtain a rather reliable calculation of the process $p + p
\rightarrow d + e^+ + \nu_e$, so important for stellar
nucleosynthesis~\cite{Schiavilla98}.  It may seem reasonable to anyone
that the latter process of weak capture of protons by protons might be
related to weak $\beta$-decay. But it is the introduction of an
isovector axial-vector hadronic current to play a role in both strong
and weak interactions which lead to the perhaps more startling
recognition of a  relation between a strong (pion production) and a
weak ($\beta$-decay) process.  We shall see how this comes about later.

Another explicit use of PCAC alone (in the form of Adler's consistency
condition) was in an envisioned re-scattering of a virtual pion from
one nucleon of a three-nucleon system.  This process establishes a
three-nucleon interaction due to two-pion exchange. Brown {\em et al.} 
showed that the three-nucleon force contribution to the binding energy
of nuclear matter could be obtained from  the isospin symmetric
pion-nucleon forward scattering amplitude extrapolated off the pion
mass shell, and was quite small~\cite{BGG68,BG69}.  This analysis
knowingly~\cite{Brown} neglected the pion-nucleon sigma term, a measure
of  chiral symmetry breaking   (the sigma term is proportional to the
non-conserved axial-vector current).  Somewhat later the full panoply
of current algebra {\em and} PCAC constraints (labelled current
algebra/PCAC) was brought to bear  on the off-shell pion-nucleon
amplitude. These current algebra/PCAC ``soft pion theorems" led to a
scenario in which the chiral symmetry breaking sigma term could not be
neglected, but instead was quite prominent in the three-body
interaction~\cite{CSB74}. However, the original insight of
Refs.~\cite{BGG68,BG69,Brown,CSB74}, based on PCAC and later on current
algebra/PCAC, that pion exchange based three-nucleon interactions are
small compared to two-nucleon interactions remains true in the
Lagrangian form of chiral symmetric theories.  A currently employed 
three-nucleon interaction according to a chiral Lagrangian is the
Brazil three-body force (TBF).  The first version of this TBF
~\cite{CDR83} had a sigma term contribution which did not come from a
Lagrangian and in a later  version~\cite{RC86} the sigma term
contribution was altered to conform to the current algebra/PCAC
constraints which had previously guided the Tucson-Melbourne two-pion
exchange TBF~\cite{TM79}.

A technical trick in Refs. \cite{BG69,CSB74}, which  directly relates
pion-nucleon scattering to a TBF contribution in nuclear matter, leads
me to my third (and final) illustrative example of chiral symmetry in
nuclear physics: pion condensation in nuclear matter.   An approximate
evaluation of a three-body diagram in an translationally invariant
system like nuclear matter can be made by summing and averaging the
active nucleon over the Fermi sea. Then one  obtains a modified
one-pion-exchange-potential between the other two nucleons, which can 
easily be evaluated in a many-body system. That is, the (now) single
exchanged pion has an effective mass $m^*$  which is proportional to
the isospin even, forward $\pi N$ amplitude multiplied by the density
(from the summation).  A useful way to think about pion condensation is
to extend the idea of a virtual pion rescattering from the active
nucleon of a three-nucleon cluster to the picture of a pion
rescattering again and again from the nucleons of nuclear matter.  The
criterion for pion condensation can be expressed in terms of $m^*$ (or
the self energy in the pion propagator) which again is directly related
to forward $\pi N$ amplitudes.  Pion condensation in neutron matter was
examined with such amplitudes subject to the current algebra/PCAC
constraints~\cite{Wilde}.  Actually,  this third example of chiral
symmetry had been discussed earlier with the aid of chiral
Lagrangians~\cite{BC81}.  This last problem has a contemporary
reverberation which is somewhat amusing in that, of the three problems
so far, it surely is the least constrained by experiment.  Yet the
relationships between the two forms of chiral symmetry have been
clarified by a small debate on, of all thing, kaon condensation in
dense nuclear matter.  This debate was between a group~\cite{YNK93}
who, fifteen years after the Tucson group~\cite{Wilde}, re-examined the
current algebra/PCAC program of pion condensation, and practitioners~\cite{TW95} of
the contemporary effective Lagrangian form of chiral symmetry known as
chiral perturbation theory.

These three examples share the idea of a virtual pion rescattering from
a nucleon (pion production in $NN$ collisions and two-pion exchange
TBFs) or from the many nucleons of nuclear matter (pion condensation).
Another example which I will not discuss much is the two-pion exchange
part of the $NN$ interaction itself.  As a Feynman diagram, this
process has  pion loops and the other three problems need only tree
diagrams.  As with the three-nucleon interaction, the first use of
chiral symmetry in the two-nucleon interaction was again by Gerry Brown who
applied the current algebra/PCAC constraints on the $\pi N$ amplitudes (and dropped
the sigma term) in a series of articles titled``Isn't it time to
calculate the nucleon-nucleon force?" and  ``Soft pioneering 
determination of the intermediate range nucleon-nucleon
interaction"~\cite{BD71}.   The chiral symmetry aspect of
two-pion exchange NN diagram can (as expected) and has been treated
with chiral Lagrangian techniques recently~\cite{Bira92,FC94}. I now
cut off   this introductory and historical survey  and turn to the
concept of ``soft pions".

Each of these  nuclear physics problems can be thought of as dependent
on a $\pi N$ scattering amplitude with at least one of the exchanged
pions off its mass shell: $q^2 = q_0^2 - \vec{q}\,^2 \neq m_{\pi}^2$. 
For example, short range correlations between two nucleons of nuclear
matter suggests that the virtual pions of a TBF  are spacelike with 
$q_0\approx 0$ and  $\vec{q}\,^2 \leq 10 m_{\pi}^2$\cite{BG69}, and the
calculation of Ref.~\cite{Wilde} found that, at the condensation
density,  $q_c^2\leq -2 m_{\pi}^2$.  The ``soft pion theorems" strictly
apply to pions with $q \rightarrow 0$ which means that every component
of the four-vector goes to zero.  In particular, since $q_0 \rightarrow
0$ then $q_0^2 = m_{\pi}^2 \rightarrow 0$ and a soft pion is a massless
pion.  In the language of QCD, this means that the quark mass goes to
zero and the chiral symmetry of the QCD Lagrangian is restored.  The
axial-vector current would be conserved if the pion was massless.  The
mass of the pion is small on the scale of the other hadrons ($
m_{\pi}^2/m_N^2 \approx 1/45)$ so one of the ideas of PCAC is that the
non-conservation of the axial vector current is small.  Another
formulation of PCAC suggests that one can make a smooth extrapolation
from the exact amplitudes with soft pions to obtain either theoretical 
amplitudes with on-mass-shell pions (``hard" pions in the old jargon)
or the off-shell amplitudes of the nuclear physics problems.  Certainly
the soft pion constraints of current algebra/PCAC are within the
assumed off-shell extrapolations used in these problems.

In the 1960's the current algebra/PCAC approach and the chiral
Lagrangian approach to chiral symmetry (and how it is broken in the
non-chiral world we do experiments in) developed in parallel and each
approach paid close attention to the other. The 1970 lectures by
Treiman on current algebra and by Jackiw on field theory provide a
useful (and pedagogical!) summary of this development\cite{cabook}. For
example, the linear $\sigma$ model was an early chiral Lagrangian 
motivated by the current algebra/PCAC program. This model reflects the feeling in the
1960's that the ultimate justification of the results obtained from a
chiral Lagrangian rests on the foundation of current algebra. On the
other hand, an early puzzle   was the current algebra demonstration
that the (observed) decay $\pi \rightarrow \gamma\gamma$ should be
zero.  In his lectures, Jackiw used the linear  $\sigma$ model to
demonstrate that the ``conventional current algebra" techniques were
inadequate. He went on from  this demonstration of a violation of the
axial-vector Ward identity with the nucleon level linear  $\sigma$
model to introduce a  study of anomalies which is documented in
Ref.~\cite{cabook}. (The quark level linear $\sigma$ model, however,
does appear to describe the decay $\pi \rightarrow \gamma\gamma$ and 22
other radiative meson decays~\cite{PVV}, so the final denouement of
this dialogue may be still to come.)  In any event, anomalies play no
role in the nuclear physics problems of these lectures, and will not be
discussed here. A very useful  pedagogical paper, specifically aimed at
the nuclear physicist, commented on the relation between the two
approaches to chiral symmetry.  In it, David Campbell showed that, in a given
chiral model field theory with a specific choice of canonical pion
fields, certain of the theorems expected from current algebra/PCAC (in particular the
Adler consistency condition) will not be true~\cite{Campbell79}. This
is one of the excellent papers which I hope the present lectures will
prepare the student to appreciate.

In 1979, Weinberg~\cite{Weinberg79} introduced a ``most general chiral
Lagrangian" constructed from powers of a chiral-covariant derivative of
a dimensionless pion field. This Lagrangian was aimed at calculating
purely pionic processes with low energy pions. The most general such
phenomenological Lagrangian, unlike earlier closed form models, is an
infinite series of such operators of higher and higher
dimensionality.   He was easily able to show that the lowest order
Feynman diagrams constructed from the Lagrangian are tree graphs.  These
tree graphs reproduce his earlier~\cite{Weinberg66} analysis of low
energy $\pi \pi$ scattering obtained from i) the Ward identities of
current algebra and ii) PCAC in the form of a smooth extrapolation from
soft to physical pions. The importance of the 1979 paper lies in its
analysis of the more complicated Feynman diagrams of the infinite
perturbation expansion of the chiral Lagrangian.  The phenomenological
Lagrangian would produce amplitudes of the form: $T \sim E^\nu$, where
$E$ is the energy. This fact was obtained using dimensional analysis
and $\nu$ is an integer determined by the structure of the Feynman
diagram. The QCD picture of chiral symmetry breaking (for example in a
world with only light $u$ and $d$ quark fields) imposes a further
constraint upon $\nu$: that more complicated diagrams necessarily have
larger values of $\nu$. Thus, provided that $E$ is smaller than some
intrinsic energy scale, $\Lambda$, the perturbation series is a
decreasing series in $E/\Lambda$.  The derivative structure of the
Lagrangian guarantees that amplitudes from loops and other products of
higher order perturbation theory produce only larger values of $\nu$. 
The Lagrangian cannot be renormalized because this is an effective
field theory where all possible terms consistent with the symmetries
assumed {\em must} be included.  The non-renormalizability means that 
more and more unknown constants appear at higher (arranged in powers of
$\nu$) and higher orders of perturbation theory but their effect is
suppressed by factors of  $E/\Lambda$.  Since it is a phenomenological
Lagrangian the unknown constants must be determined by experiment, and
one hopes that meaningful results can be obtained at a low enough
energy  such that the  number of terms needed remains tractable.  That is
the disadvantage of this approach.  An advantage is the systematic
nature of the scheme with respect to the breaking of chiral symmetry. I
quote from the seminal paper: ``the soft $\pi$ and soft K results of
current algebra, which would be precise theorems in the limit of exact
chiral symmetry, become somewhat fuzzy, depending for their
interpretation on a good deal of unsystematic guesswork about the
smoothness of extrapolations off the mass shell.  $\ldots$ 
phenomenological Lagrangians can serve as the basis of an approach to
chiral symmetry breaking, which has at least the virtue of being
entirely systematic"~\cite{Weinberg79}

The introduction of the nucleon into this scheme (now called chiral
perturbation theory or ChPT) led to a major industry in particle physics 
and to a reversal of the old idea that a symmetry imposed on an
effective Lagrangian can only be legitimized by an underlying theory
such as current algebra~\cite{compare}.  The new effective field theory
program does not attempt to find really fundamental laws of nature, but
does exploit systematically the symmetries encoded in the
phenomenological Lagrangian.  
 The belief in the power of this program leads to
astounding remarks in chiral perturbation theory papers. 
Consider:
\begin{quote} 
``Although the purpose of this comment is not to discuss the experimental
situation, it may be one of nature's follies that experiments seem to
favour the original LEG [Low Energy Guesses of pion photoproduction from
the nucleon] over the correct LET [Low Energy Theorems from ChPT].  One
plausible explanation for the seeming failure of the LET is the very
slow convergence of the expansion in $m_{\pi}$."~\cite{EM95}
\end{quote}
Or:
\begin{quote} 
 ``We can compare the situation with that of the decay $\eta \rightarrow
\pi^0 \gamma\gamma$ where the one loop ChPT prediction is approximately
170 times smaller than the experimental result.  The $ {\cal O} (p^6) $
contributions [the next order in the expansion] then bring the ChPT
result into satisfactory accord with experiment."~\cite{Maltman97}
\end{quote}
On a more cautious note:
\begin{quote}
``The one-loop calculation [of $\gamma\gamma \rightarrow \pi^0 \pi^0 $]
in ChPT disagrees with the data even near threshold." $\ldots$  ``In
conclusion, a self-consistent, quantitative description of 
$ \gamma\gamma \rightarrow \pi^0 \pi^0 $ and 
$\eta \rightarrow \pi^0 \gamma\gamma$ data at $ {\cal O} (p^6) $ is
still problematic.  A good description of the 
$ \gamma\gamma \rightarrow \pi^0 \pi^0 $ cross section has been achieved
whereas a satisfactory, quantitative prediction of the decay width seems
to be beyond the reach of an ordinary calculation at $ {\cal O} (p^6)
$ [such a calculation involves tree-level, one- and two-loop Feynman
diagrams]."~\cite{Scherer96}
\end{quote}
Finally, on elastic $\pi N$ scattering, the subject of these lectures:
\begin{quote}
``the chiral expansion converges to the experimental values, but the
convergence seems to be rather slow, in a sense that contributions to
different orders are comparable.  This fact seems to show that despite
of the relative success in describing elastic $\pi N$ scattering at
threshold, the third order is definitely not the whole story.  A
complete one-loop calculation, which will include the fourth order of the
chiral expansion, is probably needed for sufficiently reliable
description of this process"~\cite{Mojzis}
\end{quote}

Although nature seems to have pulled up its socks since the first
comment was made (more recent measurements of
pion photoproduction seem to favor ChPT results near threshold),
one is still left with a not fully satisfied feeling by these
comments.

More recently Weinberg applied this procedure to systems with more than
one nucleon~\cite{Weinberg90} so that effective field methods could be
extended to  nuclear forces and nuclei~\cite{footnote}.  This program is being
continued vigorously by van Kolck and others, thereby generating
another minor industry in nuclear physics a decade or so after ChPT hit
particle physics.

In the following lectures, I will review the current algebra and PCAC
program and its applications to the three nuclear physics problems of
this introduction.  These problems have also been attacked by the
effective field theory program.  The former approach to chiral symmetry
can be closely tied to the experimental program in pion-nucleon
scattering and the latter approach takes some of its undetermined
constants from pion-nucleon scattering.  Before proceeding, I recommend
the following review articles on this field.  Pion-nucleon scattering
is treated, more extensively than I will, in {\em Field Theory, Chiral
Symmetry, and Pion-Nucleus Interactions} by D. K.
Campbell~\cite{Campbell77}.  The mathematical aspects of global
symmetries in Lagrangian forms of field theory is discussed cogently in
lectures at an earlier Indian-Summer School: {\em Elements of Chiral
Symmetry} by M. Kirchbach~\cite{Kirchbach,Byers}.  A very useful
account (which I shall freely borrow from) of the original approach to
chiral symmetry is {\em Current algebra, PCAC, and the quark model} by
M. D. Scadron~\cite{Scadronrev}.  The nuclear physics aspects of
effective field theories are well described in {\em Effective Field Theory 
of Nuclear Forces} by U. van Kolck~\cite{Bira} and {\em
Dimensional Power Counting in Nuclei} by J. L. Friar~\cite{Friar97}. 
The titles of these review articles should suggest to the student where
to go for further studies.


\section{Kinematics}
	We begin with the scattering amplitude for 
$\pi ^j (q) + N(p) \rightarrow \pi ^i (q') + N(p')$ where $p,q,p',q'$
are nucleon and pion initial and final momenta. We ignore for
the time being spin and isospin 
 aspects of the  problem ($i$ and $j$ 
are pion (Cartesian) isospin indices).  For elastic scattering $q + p
= q' + p'$ and the scalar product of these four-vectors is $a b \equiv
a_0b_0 - \vec{a}\cdot\vec{b}$.  Define the ``$s$-channel" Mandelstam
invariants

\begin{flushleft}
\begin{picture}(100,100)(0,0)
\DashArrowLine(0,0)(35.858,35.858){4}
\Text(35,15)[]{$q,j$}
\ArrowLine(64.142,64.142)(100,100)
\Text(72,15)[]{$p$}
\DashArrowLine(35.858,64.142)(0,100){4}
\Text(35,90)[]{$q',i$}
\ArrowLine(100,0)(64.142,35.858)
\Text(75,90)[]{$p'$}
\CCirc(50,50){20}{Black}{White} 
\LongArrow(50,0)(50,10)
\Text(50,-10)[]{$s$}
\LongArrow(5,50)(15,50)
\Text(0,50)[]{$t$}
\end{picture}
\end{flushleft}

\begin{flushright}
\vspace{-100pt}
\begin{displaymath}
\begin{array}{ccccl}
s & \equiv & (p+q)^2 & = & (p' + q')^2  \\ 
t & \equiv & (q-q')^2 & = & (p' -p)^2  \\ 
u & \equiv & (p-q')^2 & = & (p' -q)^2\;. 
\end{array}
\end{displaymath}
\end{flushright}

\vspace{50pt}

The invariant $s$ in this $s$-channel corresponds to the square of the
total energy for the process.  Since four-momentum conservation is but
one constraint upon the four momenta, there are three independent
combinations of these momenta (and energies), but only two independent
combinations of Lorentz scalar products. So $s$, $t$, and $u$ are not
independent and it can quickly be shown with the aid of $q + p= q' + p'$ 
that 
\[ s + t + u = p^2 + p'^2 + q^2 + q'^2\;.  \]
If the nucleons are on-mass-shell ($p^2 = p^2_0 - \vec{p}\,^2 = m^2$)
and the pions are on-mass-shell ($q^2 = q^2_0 - \vec{q}\,^2 = \mu^2$),
this relation becomes $s + t + u = 2m^2 + 2\mu^2$.

These Lorentz invariants $s$, $t$, and $u$ can be visualized 
in different coordinate systems.  For example, in the $s$-channel
center of mass frame the incoming (on-mass-shell) momenta are pion $q=
(q_0,\vec{q}_{cm})$  and nucleon $p=(E,-\vec{q}_{cm})$,  the final
(on-mass-shell) momenta are  $q'= (q_0,\vec{q}_{cm}\!\!^{\prime})$ and
$p'=(E,-\vec{q}_{cm}\!\!^{\prime})$, where $|\vec{q}_{cm}| =
|\vec{q}_{cm}\!\!'|$ and the three-vector $\vec{q}_{cm}$ is simply
rotated by the angle $\theta_{cm}$.

\begin{center}
\begin{picture}(350,100)(0,0)
\DashArrowLine(0,50)(70.711,50){4}
\ArrowLine(141.42,50)(70.711,50)
\Vertex(70.711,50){1}
\LongArrow(175,50)(200,50)
\Vertex(250,50){1}
\DashArrowLine(250,50)(300,100){4}
\ArrowLine(250,50)(200,0)
\DashLine(250,50)(300,50){3}
\ArrowArc(250,50)(25,0,45)
\Text(285,60)[]{$\theta_{cm}$}
\Text(30,35)[]{$q= (q_0,\vec{q}_{cm})$}
\Text(115,35)[]{$p=(E,-\vec{q}_{cm})$}
\Text(235,25)[lc]{$p^{\prime}=(E,-\vec{q}_{cm}\!\!^{\prime})$}
\Text(285,75)[lc]{$q^{\prime}= (q_0,\vec{q}_{cm}\!\!^{\prime})$}
\end{picture}
\end{center}
In this frame one can evaluate
\begin{displaymath}
\begin{array}{ccccl}
 s  & \equiv & (p+q)^2 & = & m^2 + \mu^2 +
 2[(m^2 + q^2_{cm})^{\textstyle{\frac{1}{2}}}
(\mu^2 + q^2_{cm})^{\textstyle{\frac{1}{2}}} + q^2_{cm} ]  \\
 t & \equiv & (q-q')^2 & = & -2 q^2_{cm}(1 - \cos \theta_{cm})  \\
 u  & \equiv & (p-q')^2 & = & m^2 + \mu^2 -
    2[(m^2 + q^2_{cm})^{\textstyle{\frac{1}{2}}}
    (\mu^2 + q^2_{cm})^{\textstyle{\frac{1}{2}}} 
    + q^2_{cm}\cos \theta_{cm}  ]\;\;.
\end{array}
\end{displaymath}   
Note that $s \geq (m + \mu)^2$ and
$t\leq 0$ for physical $\pi N$ scattering where $q^2_{cm} \geq
0$ and $-1 \leq \cos \theta_{cm} \leq 1$.  The $cm$ energy of the
on-mass-shell nucleon is $ E = (s-m^2 - \mu^2)/2\sqrt{s} $.  Partial
wave phase shifts are naturally expressed in terms of $q^2_{cm}$ and its
associated Mandelstam variable $s$.

Now consider the $s$-channel laboratory frame in which the target
nucleon is at rest and the kinetic energy of the incoming pion is
defined by $T_{\pi}\equiv \omega - \mu = \sqrt{k^2 + \mu^2} -\mu$, where
$\omega$ is the lab energy of the incoming pion:

\begin{center}
\begin{picture}(350,100)(0,0)
\DashArrowLine(0,50)(70.711,50){4}
\Vertex(70.711,50){2}
\LongArrow(175,50)(200,50)
\Vertex(250,50){2}
\DashArrowLine(250,50)(300,100){4}
\ArrowLine(250,50)(300,0)
\DashLine(250,50)(300,50){3}
\ArrowArc(250,50)(25,0,45)
\Text(285,60)[]{$\theta_{lab}$}
\Text(30,35)[]{$q= (\omega,\vec{k})$}
\Text(115,35)[]{$p=(m,0)$}
\Text(285,25)[lc]{$p^{\prime}=(E,\vec{p})$}
\Text(285,75)[lc]{$q^{\prime}= (\omega',\vec{k}^{\prime})$}
\end{picture}
\end{center}
In this laboratory frame the Mandelstam invariants $s$ and $u$ take the form
\begin{displaymath}
\begin{array}{cccclll}
 s & \equiv & (p+q)^2 &= &  m^2 + \mu ^2 + 2m \omega &=&
 		 (m + \mu)^2 + 2m T_{\pi} \\
   u & \equiv & (p-q')^2 &=&  m^2 + \mu ^2 - 2m \omega' & .      
\end{array}
\end{displaymath}  
In either frame, it is clear that the threshold for physical $\pi N$
scattering is $s_{th} = (m + \mu)^2$ from $q^2_{cm}=0$ or $T_{\pi} =0$,
$t_{th} = 0$ from $q^2_{cm}=0$,
and $u_{th} = (m - \mu)^2$ from $q^2_{cm}=0$ or $\omega' = \mu$. 

 Now
introduce the variable
\[    \nu = \frac{s -u}{4m}\;,   \]
which has the threshold value $\nu_{th} = \mu$ in the $s$-channel.  For
on-shell nucleons and pions $\nu = \omega + t/(4m)$ so that in the
forward direction, $ t = 0$, the variable $\nu$ represents the lab
energy of the incoming pion.  Pion-nucleon scattering amplitudes are
often given in terms of the pair of variables
$(\nu,t)$ rather than $(s,t)$ which would be appropriate for a partial
wave representation, for example.  A reason for this is that the
variable $\nu$ has a definite symmetry under {\em crossing}, a concept to
which we now turn.  Crossing is the interchange of a particle with its
antiparticle with opposite four-momentum.  I can ``cross" the pions
by adding nothing to
 the $s$-channel relation 
$\pi ^j (q) + N(p) \rightarrow \pi ^i (q') + N(p')$ as follows:

\begin{displaymath}
\begin{array}{ccccccc}
\pi ^j (q) &          &+ N(p) &\rightarrow &\pi ^i (q') &  &+ N(p') \\
           &\overbrace{\pi^i(q') + \bar{\pi}^i (-q')} & & & &
            \overbrace{\pi^j(q) + \bar{\pi}^j (-q)} &  \\
\end{array}
\end{displaymath} 

Since I have changed nothing this $s$-channel process is equivalent to 
$\bar{\pi}^i (-q') + N(p) \rightarrow \bar{\pi}^j (-q) + N(p')$:
\begin{flushleft}
\begin{picture}(100,100)(0,0)
\DashArrowLine(35.858,35.858)(0,0){4}
\Text(35,15)[]{$-q',i$}
\ArrowLine(64.142,64.142)(100,100)
\Text(72,15)[]{$p$}
\DashArrowLine(0,100)(35.858,64.142){4}
\Text(35,90)[]{$-q,j$}
\ArrowLine(100,0)(64.142,35.858)
\Text(75,90)[]{$p'$}
\CCirc(50,50){20}{Black}{White} 
\LongArrow(55,0)(55,10)
\Text(50,-10)[]{$u-{\rm channel}$}
\end{picture}
\end{flushleft}
\vspace{5pt}
which is called the $u$-channel because $u = (p-q')^2$ is now the sum of
the incoming momenta and in this channel $u$ is the square of the total
energy.  Because the antiparticle of a pion is still a pion 
($\overline{\pi^+} = \pi^-$ and $\overline{\pi^0} = \pi^0$)
both the 
$s$-channel and the $u$-channel describe $\pi N$ scattering.

Carrying on with ``crossing" one can convince oneself that
\begin{description}
	\item [$s$-channel] In this channel $s$ is the total energy
	squared and for physical scattering $s \geq s_{th} = (m+\mu)^2$
	and $t\leq 0$  (and $u\leq 0$).
	
	\item [$u$-channel] In this channel $u$ is the total energy
	squared and for physical scattering $u \geq u_{th} = (m+\mu)^2$
	and $t\leq 0$  (and $s\leq 0$).

	\item [$t$-channel] In this channel the incoming particles are a
	pion and an anti-pion and the outgoing particles are $N$ and
	$\bar{N}$.  For physical scattering the total energy squared
	must be larger than the rest mass of the heaviest particles, so
	that $t \geq t_{th} = (m +m)^2$, $s\leq 0$, and $u\leq 0$.
\end{description}
The scattering amplitude $T(s,t,u)$ is a function of the three (not all
independent) variables.   The physical regions of the variables of the
three channels are disjunct.  We have determined the threshold values
``by inspection".  It is slightly more complicated to work out the
 boundaries of the physical regions in the
Mandelstam plane.They are given by the zeros of the Kibble
function\cite{Kibble}
\[  \Phi = t[su - (m^2 - \mu^2)]\,.   \]
The physical regions correspond to the regions where $\Phi \geq 0$.
This criterion essentially characterizes the need for the scattering
angle to satisfy $-1 \leq \cos \theta_{cm} \leq 1$.  Clearly $t=0$ or
$\cos\theta_{cm} = 1$ is a boundary of the physical region no matter
what the values of $s$ and $u$ are. The other zero of $\Phi$ then shows
the dependence of a lower limit to $t$ for $s$-channel $\pi N$
scattering (for example) which
depends upon the values of $s \geq s_{th} = (m+\mu)^2$ and  $u\leq 0$. 
 Elastic scattering depends upon two independent variables:
some sort of energy and some sort of scattering angle.  As mentioned
before, if you wanted to end up with a partial wave representation the
natural variables are the pair ($s,t$) because $t$ has a simple
interpretation in terms of $q^2_{cm}$ and $\theta_{cm}$, for example.  In the
following discussion,  the pair ($\nu,t$) is  more natural because $\nu
\rightarrow -\nu$ under the interchange of $s$ and $u$: $s
\rightleftharpoons u$.  Then the physical regions of the Mandelstam plane
are bounded by a hyperbola in the $(\nu,t)$ plane and the straight line 
$t=0$ (See Fig 1).  The boundaries of the physical regions for 
$\pi + N \rightarrow \pi + N$ and for $\pi + \pi  \rightarrow N +
\bar{N}$ form branches of the same hyperbola.
Note that the asymptotes of the boundary hyperbola
are the lines $s=0$ and $u=0$.


\begin{figure}[htpb]
\unitlength1.cm
\begin{picture}(6,13)(0,3.5)
\includegraphics{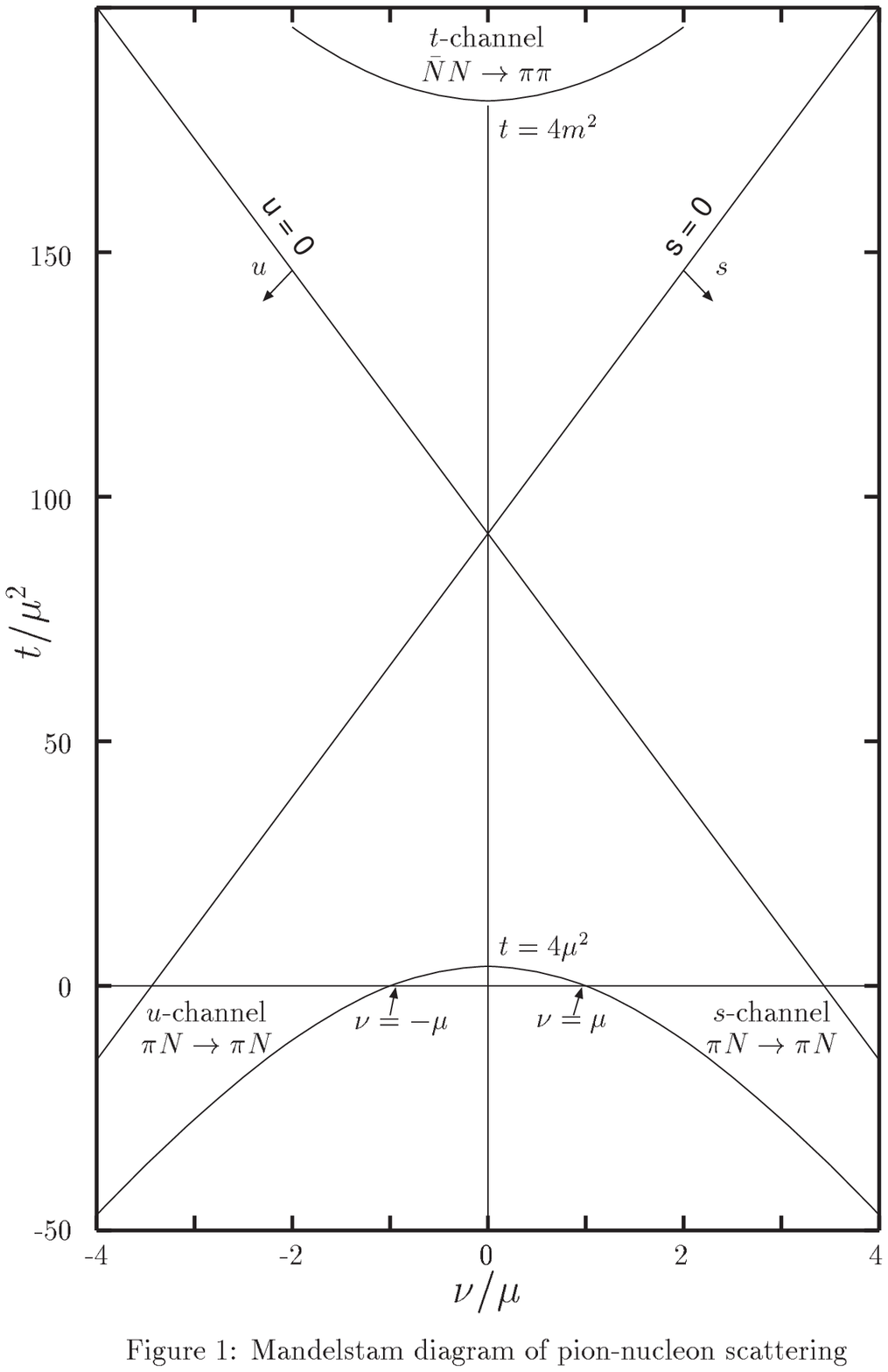}
\end{picture}
\end{figure}

Let us turn from the relativistic invariants in $T(\nu,t)$ to 
isospin considerations in the three channels.   The incoming
particles in the $s$ and $u$ channels  have isospin 
${\textstyle{\frac{1}{2}}}$ (nucleon) and $1$ (pion).  The total isospin
$I_s$ is then either ${\textstyle{\frac{1}{2}}}$ or
${\textstyle{\frac{3}{2}}}$.  In the $t$-channel
($\pi\bar{\pi}\rightarrow N \bar{N}$) $I_t = 0,1$ so there are also two
amplitudes in isospin: $T^{(+)}$ with $I_t = 0$ and $T^{(-)}$
 which has $I_t = 1$.
 The $t$-channel isospin is especially convenient because the pions
obey Bose symmetry when crossed:  ie $I^0_t \rightarrow I^0_t$ and
$I^1_t \rightarrow -I^1_t$ when $s\rightleftharpoons u$.  To make contact
with the $s$-channel amplitudes (and the charge states) we need the  
$s\rightleftharpoons t$ crossing relations: 

\begin{displaymath}
\begin{array}{cclcccl}
 T^{(+)} &=& {\textstyle{\frac{1}{3}}} (T^{\textstyle({\frac{1}{2}})} +
 		2T^{\textstyle({\frac{3}{2}})} ) & \hspace{20pt}&
T^{(-)} &=& {\textstyle{\frac{1}{3}}} (T^{\textstyle({\frac{1}{2}})}
		-T^{\textstyle({\frac{3}{2}})}) \\
T^{\textstyle{(\frac{1}{2}})} &=&  T^{(+)} +2 T^{(-)} &\hspace{20pt}&
T^{\textstyle{(\frac{3}{2}})} &=& T^{(+)} - T^{(-)}\;.
\end{array}
\end{displaymath}
The pion field operators transform as components of a vector in isospin
space with  Cartesian components defined as~\cite{phasefootnote}

\begin{eqnarray*}
	\pi^+ &=& +(\pi^1 + i \pi^2)/\sqrt{2}  \\
	\pi^0 &=& \pi^3   \\
	\pi^- &=& +(\pi^1 - i \pi^2)/\sqrt{2}\;,  
\end {eqnarray*}
and 
\begin{displaymath}
\begin{array}{ccccc}
	T(\pi^+p \rightarrow \pi^+p) &=& T^{\textstyle({\frac{3}{2}})}
		&= & T^{(+)} - T^{(-)}\\
	T(\pi^-p \rightarrow \pi^-p) &=&{\textstyle{\frac{1}{3}}}
 	T^{\textstyle({\frac{3}{2}})} + {\textstyle{\frac{2}{3}}}
	T^{\textstyle({\frac{1}{2}})} &=& T^{(+)} + T^{(-)}\\
	T(\pi^-p \rightarrow \pi^0 n) &=& {\textstyle{\frac{\sqrt{2}}{3}}}
	(T^{\textstyle({\frac{3}{2}})} - T^{\textstyle({\frac{1}{2}})})
	&=& -\sqrt{2}T^{(-)}\;,
\end{array}
\end{displaymath}	
for example.

Now we are in a position to examine the isospin structure of the
T-matrix elements  which describe the scattering 	
$\pi ^j (q) + N(p) \rightarrow \pi ^i (q') + N(p')$.  They are defined as:
\begin{equation}
\langle q' p'|S - 1| q p \rangle \equiv +i (2\pi)^4 
	\delta^4 (p' + q' -p -q) T^{ij} (\nu,t; p^2=m^2, p'^2=m^2, q^2,
	q'^2)\;,	\label{eq:tmat}
\end{equation}	
where we have displayed the (assumed) on-mass-shell nucleons and left
the four-momentum of the pions as a variable.	The isospin structure of
$T^{ij}$ is perhaps most easily visualized from the Feynman diagram with
 an intermediate nucleon pole state and the isospin ``scalar" vertex
 $\bar{N}\vec{\tau} N \cdot \vec{\pi}$:
\begin{eqnarray}
 T^{ij} \tau^i\tau^j &=& T^{(+)}{\textstyle{\frac{1}{2}}}
	(\tau^i\tau^j +\tau^j\tau^i) +
	T^{(-)}{\textstyle{\frac{1}{2}}}(\tau^i\tau^j -\tau^j\tau^i)
	\nonumber \\   \label{eq:isospin}
	&=& T^{(+)}\delta^{ij} + T^{(-)} i \epsilon^{ijk} \tau^k\;.
\end{eqnarray}	
The second equality is easily proved from the  properties of
the 	 $SU(2)$ $\tau$-matrices: $\{ \tau^i,\tau^j \} = 2 \delta^{ij}$
and	 $[ \tau^i,\tau^j ] = 2i \epsilon^{ijk} \tau^k$.  With this
representation, it is clear that $T^{(+)}$
($T^{(-)}$) must be even (odd) under the interchange of the pions
$i\leftrightarrow j$ and 	 $s \leftrightarrow u$.   We also note
that $\pi^i \pi^j \delta^{ij} = \vec{\pi}\cdot \vec{\pi} = \pi^2$ could
be realized by the $t$-channel exchange of an isoscalar scalar $\pi\pi$
resonance--the putative sigma meson.  In a similar manner, the 
$t$-channel odd exchange, $\pi^i \pi^j i \epsilon^{ijk} \tau^k = i
\vec{\pi} \times \vec{\pi}\cdot \vec{\tau}$ could be realized by the
isovector $\rho$ meson.  We will discuss these models in a later lecture.

Finally, we let the Dirac spinors $u(p)$ carry the spin of the nucleons
and write the Lorentz invariant $T = +\bar{u}(p')\{
M(\nu,t)\}u(p)$ where $M$ could be made up of scalars, vectors, and
higher order tensors constructed of the vectors $p,p',q,q'$ and the
gamma matrices  $1,\gamma^\mu, \gamma^\mu\gamma^\nu, \gamma_5,
\gamma^\mu \gamma_5$.  That is, one could form 
\begin{displaymath} {M = A + B^\mu
\gamma_\mu + C^{\mu\nu} [\gamma_\mu,\gamma_\nu] +
D^\mu\gamma_\mu\gamma_5 + E\gamma_5}
\end{displaymath}
but conservation of parity
eliminates $D^\mu$ and $E$ as candidates.  With the aid of the Dirac
equation for free (on-mass-shell) nucleons, 
\[ (p^\mu\gamma_\mu - m) u(p)
= 0 = \bar{u} (p') (p'^\mu\gamma_\mu - m)\;, \]
all the combinations one can
write down for $C^{\mu\nu}$ reduce to $A + B^\mu \gamma_\mu$  where $A$
is a scalar and $B$ is a four-vector formed of those available:
$p,p',q,q'$. $ B^\mu$ cannot be $p$ or $p'$ because the Dirac equation
would make $T \sim m\bar{u}(p')u(p)$ already included in $A$.  So $ B^\mu$
must be linear in $q$ and $q'$, but $ B^\mu$ cannot be $(q-q')^\mu = 
(p'-p)^\mu$ for the same reason. We conclude that
\begin{displaymath}
	 T^{\pm} = \bar{u}(p')\{A^{\pm}(\nu,t) +{\textstyle{\frac{1}{2}}}
	(\rlap/q' + \rlap/q) B^{\pm}(\nu,t)\}u(p)   
\end{displaymath}
where $\rlap/q \equiv q^{\mu}\gamma_\mu$ and
the factor ${\textstyle{\frac{1}{2}}}$ is inserted to make the
expressions for $s$ and $u$-channel nucleon poles in $B$ simple.
With the aid of $\nu = (s-u)/4m = (q' + q)\cdot(p' + p)/4m$ and the free
particle Dirac equation, one can rewrite this as
\begin{equation} T^{\pm} = \bar{u}(p')\{A^{\pm}(\nu,t) + \nu B^{\pm}(\nu,t) 
	- {\textstyle{\frac{1}{4m}}}[\rlap/q,\rlap/q']B^{\pm}(\nu,t))\} u(p)\;.
	\label{eq:FandB} 
\end{equation}
Define the combination $A + \nu B =F$, which is called $D$ in
H\"{o}hlers book~\cite{hohlerbook} and in much of the literature.  It
can be shown that this combination of invariant amplitudes corresponds
to the  non-relativistic forward ($p = p'$) scattering of a nucleon from
a pion in which the spin of the nucleon remains unchanged (non-spin
flip); for example, see Ref. \cite{Campbell77}, pp 612.  For this reason
the invariant amplitude $F$ is sometimes called the ``forward amplitude"
but obviously we can study the combination $A + \nu B$ for any value of
$\nu$ and $t$.

Expressions of chiral symmetry in the form of soft pion theorems and
their  on-mass-shell analogues are most naturally expressed as
conditions on the  four amplitudes  $F^{\pm}(\nu,t)$ and
$B^{\pm}(\nu,t)$,  rather than the set $A^{\pm}(\nu,t)$ and
$B^{\pm}(\nu,t)$.  As we shall see in the following, the predictions of
chiral symmetry breaking are all in the subthreshold crescent of the
Mandelstam representation (Figure 2).   This crescent is below the $s$-channel
threshold $s_{th} = (m + \mu)^2$ for $\pi N \rightarrow  \pi N$, below
the $u$-channel threshold $u_{th} = (m + \mu)^2$ for 
$\bar{\pi} N \rightarrow  \bar{\pi} N$, and below the
$t$-channel threshold $t_{th} = (m + m)^2$ for $\bar{\pi} \pi \rightarrow 
\bar{N} N $.  Therefore the invariant amplitudes in this subthreshold
crescent are real functions of the real variables $\nu$ and $t$.

\begin{figure}
\unitlength1.cm
\begin{picture}(6,9)(2,11.5)
\includegraphics{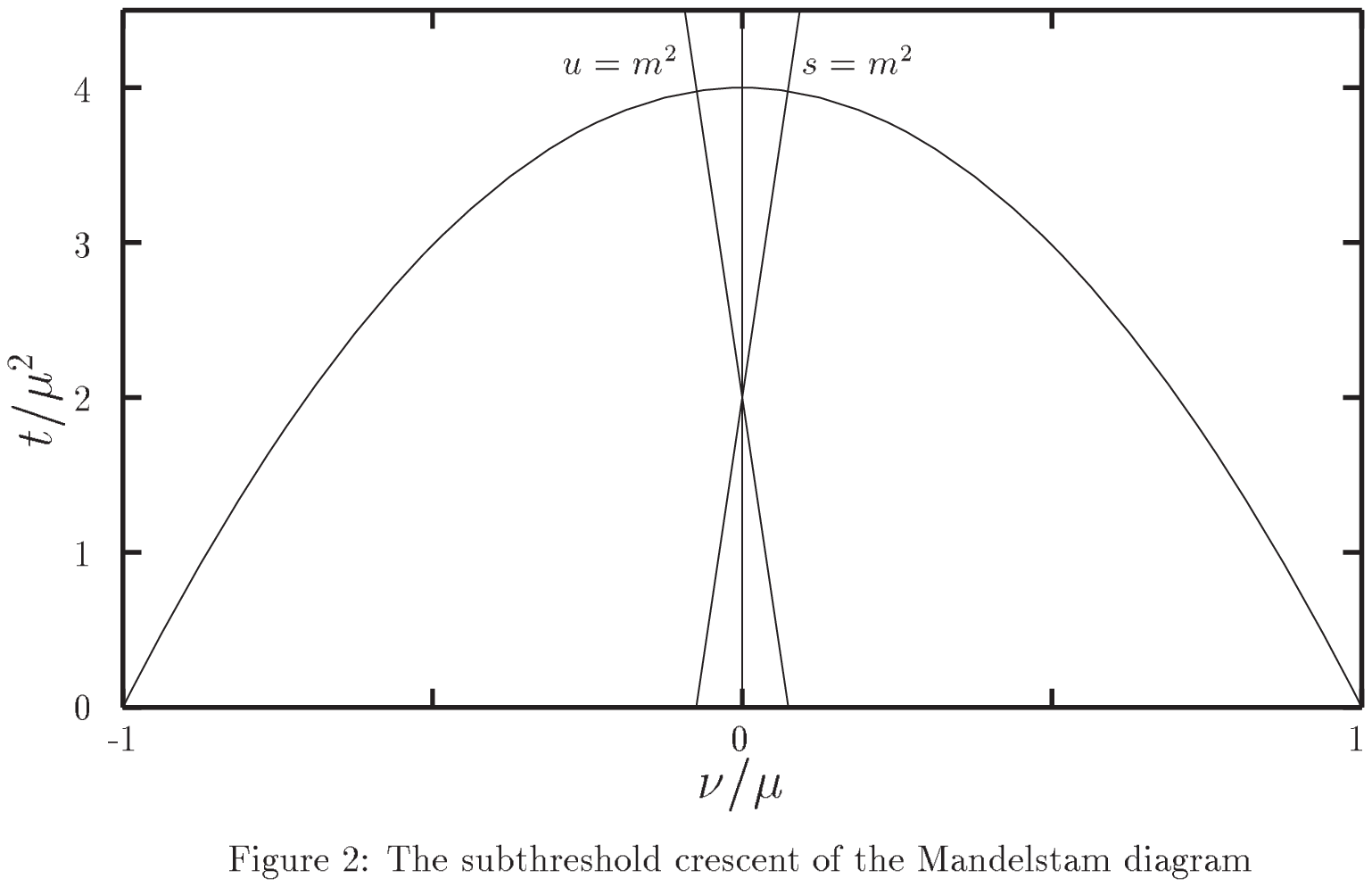}
\end{picture}
\end{figure}

\section{Current Algebra and PCAC}

The formalism in this lecture follows very closely the exposition of
Scadron in his review article~\cite{Scadronrev} and
textbook~\cite{Scadronbook}.  It is included here to make the lectures
somewhat self-contained and to enable me to compare the current algebra
predictions  for pion-nucleon scattering with the current experimental
results, a comparison  which has not been emphasized in most contemporary
discussions of the meson factory data.

We begin with the basic ideas of the current algebra-Partially Conserved
Axial-vector Current (PCAC) implementation of chiral symmetry in hadronic
physics.  Recall that in non-relativistic quantum mechanics the charge
operator $Q(t)$ obeys the Heisenberg equation of motion:
\begin{equation}
	\frac{dQ(t)}{dt} = \frac{\partial Q(t)}{\partial t}
	- i [Q(t),H(t)] \;.  \label{eq:heq}
\end{equation}
Then, if $Q$ is explicitly independent of time, conservation of charge
is equivalent to $Q$ commuting with the Hamiltonian.  In relativistic
quantum mechanics we can define current densities and Hamilton densities 
as
\begin{eqnarray}
	Q(t) & \equiv & \int d^3x J_0(t,\vec{x})\\
	H(t) & \equiv & \int d^3x {\cal H}(t,\vec{x})\;.
\end{eqnarray}
The equation of continuity for charge
\begin{equation}
	\frac{dQ(t)}{dt} = \int d^3x \left(\frac{\partial J_0(t,\vec{x})} 
	{\partial t}\right) + \vec{\nabla}\cdot \vec{J}(t,\vec{x})
	\equiv \int d^3x \partial J(x) \label{eq:eqofcon}
\end{equation}
allows one to rewrite (\ref{eq:heq}) in the local density form
\begin{equation}
     	i \partial J(x) = [Q, {\cal H}(x)]    \label{eq:heq2}
\end{equation}
if $Q$ does not depend explicitly on time.

Commutators such as this one form the underlying dynamics in current
algebra. We now know that at the hadronic level QCD is spontaneously
broken  down into a vector $SU(2)$ algebra and an axial-vector $SU(2)$
algebra.  Current algebra is based on the $SU(2)$
equal-time commutation relations of isotopic vector charges
\begin{eqnarray}
		[Q^i,Q^j] &=& i \epsilon^{ijk}Q^k    \label{eq:QQ}
\end{eqnarray}
which was extended in the 1960's by Gell-Man's suggestion of adding
axial charge ($Q^i_5$) commutators
\begin{equation}
	[Q^i,Q^j_5]  =  i \epsilon^{ijk}Q^k_5\;,  \hspace{20pt}
	[Q^i_5,Q^j_5]  =  i \epsilon^{ijk}Q^k    \label{eq:QQ5}
\end{equation}
to complete the chiral algebra.  Models of $\cal H$ for strong,
electromagnetic, and weak transitions as products of currents then
predict observable hadron current divergences according to 
(\ref{eq:heq2}) with the aid of the charge algebra and its
current algebra generalizations.  We defer discussion of the current algebra
per se  until after this introductory material is discussed.

The $SU(2)$ notation is the same as before with states $|\pi^i\rangle$,
and  (to be defined) vector currents $J^i$ and axial-vector currents 
$A^i$, where $i=1,2,3$.  Define isotopic charges from current densities
as $Q^i = \int d^3x J_0^i(t,\vec{x})$.  The $SU(2)$ hadron states
transform irreducibly as $Q^i|P^j\rangle = i f^{ijk}|P^k\rangle$, where
$f^{ijk} = \epsilon^{ijk}$.  In the  generalization
to $SU(3)$ the anti-symmetric structure constant $f^{ijk}$ is related
to the Gell-Man $\lambda^i$ matrices, $i = 1,\cdots8$ (for a tabulation,
see Ref.~\cite{Scadronrev}).  Now consider the
$SU(2)$ and $SU(3)$ structure of the electromagnetic current
\begin{equation}
	J^\gamma_\mu = J^S_\mu + J^V_\mu = 
	{\textstyle{\frac{1}{\sqrt{3}}}}J^8_\mu + J^3_\mu\;,
\end{equation}
where $J^V_\mu = J^3_\mu$ is the isovector current and $J^S_\mu =
{\textstyle{\frac{1}{\sqrt{3}}}}J^8_\mu$ corresponds to $2J^Y_\mu$ the
hypercharge current.  The corresponding charges are
\begin{equation}
	Q = \int d^3x J_0^\gamma(x),\hspace{20pt}
	{\textstyle{\frac{1}{2}}}Y = \int d^3x J_0^S(x), \hspace{20pt}
	I_3 = \int d^3x J_0^V(x)\; .
\end{equation}
The equation of continuity (\ref{eq:eqofcon}), coupled with the fact
that the electromagnetic charge $Q$ is conserved in the strong interaction,
implies $\partial J^\gamma (x) = 0$.  The $SU(3)$ structure of the
photon is consistent with the separate conservation of isospin and hypercharge in
the strong interactions and suggests the Gell-Mann-Nishijima relation $Q = 
{\textstyle{\frac{1}{2}}}Y + I_3$.

\subsection{Conserved $SU(2)$ Vector Currents}

We want to treat $J^S_\mu$ and $J^{V,i}_\mu$ (where $J^V_\mu \equiv
J^{V,i}_\mu$) as conserved hadronic currents for the strong interactions,
$\partial J^{V,i} =0$ and $\partial J^S =0$.
To illustrate this, define the isovector part of the $SU(2)$ strong vector
current by its nucleon matrix elements:
\begin{equation}
  \langle N_{p'}|J^i_\mu(x)|N_p\rangle =
  \bar{N}_{p'}{\textstyle{\frac{\tau^i}{2}}}
      [F^V_1(q^2) \gamma_\mu + F^V_2(q^2)i \sigma_{\mu\nu} q^\nu
      /2m ] N_p e^{iq\cdot x}\;,   \label{eq:isovec}
\end{equation}
where ($q = p' -p$). $F^V_1(q^2)$ is the nucleon isovector charge form factor and 
$F^V_2(q^2)$ the nucleon isovector magnetic moment form factor.  The
isoscalar and isovector decomposition is defined as $F^S_{1,2}(q^2)=
F^p_{1,2}(q^2) + F^n_{1,2}(q^2)$ and $F^V_{1,2}(q^2)= F^p_{1,2}(q^2) -
F^n_{1,2}(q^2)$.  This definition is made because we {\em identify} the
$i=3$ component of $J^V_\mu$ plus $J^S_\mu$ (with a similar definition)
as the electromagnetic current for the proton or the neutron.   For the
complete electromagnetic current, $J^\gamma_\mu = J^S_\mu + J^V_\mu$,
charge is conserved and $F^V_1(0) = 1$.  In terms of isospin this
conservation law becomes  $F^V_1(0) = F^p_1(0) + F^n_1(0) = 1$, where 
$ F^p_1(0) =1$ and $F^n_1(0) = 0$.  With the aid of the free Dirac
equation one can show $\bar{u}_{p'}q^\mu \gamma_\mu u_p = 0$ and 
$q^\mu \sigma_{\mu\nu} q^\nu = 0$, thus demonstrating that the
divergence of the  isovector  current (\ref{eq:isovec}) (and the analogue
isoscalar current) is indeed zero.

In a similar manner, we can extend the electromagnetic charged pion
current to the isovector-vector hadron current:
\begin{equation}
   \langle \pi^i_{p'}|J^{V,j}_\mu(x)|\pi^k_{p}\rangle  =
	\epsilon^{ijk} F_\pi(q^2) (p' + p)_\mu e^{iq\cdot x}\;,
	\label{eq:isovecpi}
\end{equation}
where the charge form factor of the pion is normalized to $F_\pi(0)=1$.
In our isospin convention $J^\gamma_\mu = J^{V,3}_\mu$, and I note that 
$J^S_\mu$ does not couple to pions; this would violate $G$-parity. 
The current of (\ref{eq:isovecpi}) is conserved for $p'\,^2 = p^2$ up to
a term proportional to $(p' -p)^\mu$ which disappears in 
$\langle \pi|\partial J^{V,j}|\pi\rangle = 0$.  But the general $SU(2)$
vector current is conserved {\em as an operator}
\begin{equation}
	\partial J^i(x) = 0  \hspace{20pt} i =1,2,3   \label{eq:opeq}
\end{equation}
such that the nucleon and pion matrix elements of (\ref{eq:opeq})
vanish, consistent with the vanishing divergences of (\ref{eq:isovec})
and (\ref{eq:isovecpi}) for on-shell equal-mass hadrons.

One can continue to demonstrate the vanishing divergence of other
matrix elements of the hadronic vector current.  For example, the
existence of the vector mesons $\rho$ and $\omega$ suggests a direct
$\rho\mbox{-}\gamma$ and $\omega\mbox{-}\gamma$ transition.  We write the
$\rho$-to-vacuum matrix elements of the hadronic isovector vector
current as 
\begin{equation}
	\langle 0|J^{V,i}_\mu(x)|\rho^j(q)\rangle =
	\frac{m^2_\rho}{g_\rho}\epsilon_\mu(q)\delta^{ij} e^{-iq\cdot x}
	\;,
\end{equation}
and the hadronic isoscalar vector current as
\begin{equation}
	\langle 0|J^S_\mu(x)|\omega(q)\rangle =
	\frac{m^2_\omega}{g_\omega}\epsilon_\mu(q)
	e^{-iq\cdot x}
	\;.
\end{equation}
These currents are conserved as well because 
$\partial J^{V,S} \propto q \cdot \epsilon(q) =0$ for on-shell spin-1
polarization vectors $\epsilon_\mu(q)$.

\subsection{$SU(2)$ Axial-vector Current $A^i_\mu$}

We introduce this current with the simplest matrix element (and the
analogue  of the $\rho$-to-vacuum matrix element of the vector current)
$\pi$-to-vacuum:
\begin{equation}
\langle 0|A^i_\mu(x)|\pi^j(q) \rangle = i f_\pi q_\mu \delta^{ij}
	e^{-iq\cdot x}  \;, \label{eq:oApi}
\end{equation}
where $f_\pi \approx 93$ MeV is called the pion decay constant and its
value is measured in the weak decay $\pi^+ \rightarrow \mu^+ \nu_\mu$.  The divergence
of (\ref{eq:oApi}) is
\begin{equation}
	\langle 0|\partial A^i(0)|\pi^j\rangle = \delta^{ij}f_\pi \mu^2
\end{equation}
for $i,j = 1,2,3$ and an on-shell pion $q^2 = \mu^2 \equiv m^2_\pi$.
From this exercise we learn that {\em axial-currents are not conserved,
even if $SU(2)$ is an exact symmetry}.  But the pion mass is small
relative to  all other hadrons: $\mu^2/m^2 \approx 1/45$.  In 1960 Nambu
suggested that $\langle 0|\partial A^i(0)|\pi^j\rangle \approx 0$ and
even $\partial A^i \approx 0$ in an operator sense~\cite{Nambu}.
Next we define the nucleon matrix elements of $A^i_\mu$:
\begin{equation}
  \langle N_{p'}|A^i_\mu(x)|N_p\rangle =
  \bar{N}_{p'}{\textstyle{\frac{\tau^i}{2}}}
      [g_A(q^2)(t) i\gamma_\mu\gamma_5 + h_A(q^2) i q_\mu\gamma_5
      ] N_p e^{iq\cdot x}\;,   \label{eq:isoax}
\end{equation}
where $q=(p' -p)$ as usual and $\bar{\gamma}_5 = \gamma_5$ as 
in Refs.\cite{Scadronrev,Scadronbook}. Finally we present a diagrammatic
representation of these matrix elements:
\begin{center}
\begin{picture}(300,100)(0,0)
\Vertex(50,50){3}
\Photon(10,50)(50,50){3}{3}
\DashArrowLine(90,50)(50,50){4}
\Text(65,58)[b]{$\pi$}
\Text(30,58)[b]{$A$}
\Text(95,50)[]{$q$}
\PText(21,51)(45)[]{>}
\Text(50,20)[]{$\langle 0|A|\pi(q) \rangle$}

\SetOffset(200,0)
\ArrowLine(64.142,64.142)(100,100)
\Text(72,15)[]{$p$}
\ArrowLine(100,0)(64.142,35.858)
\Text(75,90)[]{$p'$}
\CCirc(50,50){20}{Black}{White} 
\Photon(-10,50)(30,50){3}{3}
\Text(10,55)[b]{$A$}
\PText(1,51)(45)[]{>}
\Text(20,10)[]{$\langle N_{p'}|A_\mu(x)|N_p\rangle$}

\end{picture}
 \\ \mbox{} \\ {Figure 3: Matrix elements of the axial-vector current}
\end{center}
which will be useful in the discussion of PCAC and later on of current
algebra.

\subsection{PCAC}
We now review three ways of looking at the partial conservation of the
axial vector current (PCAC) and establish a {\em soft-pion theorem}
which will be used and tested against data in the following.  The first
(Nambu) statement of PCAC is simply that  $\partial A^i \approx 0$ in
an operator sense. We now consider the general emission of a very
low energy pion $A \rightarrow B + \pi^i$ so that the emitted pion is
soft ($m_{\pi}^2 \approx 0$).  Replace the pion by an axial-vector and
then remove the pion pole in this diagrammatic way:

\begin{figure}[htbp]
\unitlength1.cm
\begin{picture}(6,3.5)(2.5,16)
\includegraphics{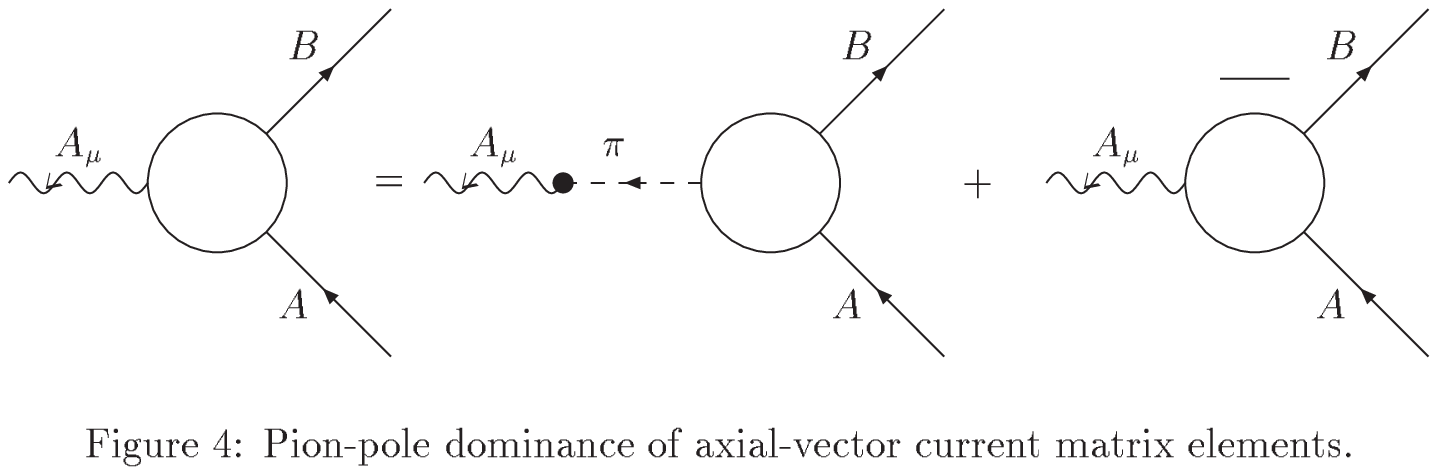}
\end{picture}
\end{figure}

This diagrammatic equation relates the axial-vector $M$-function $M_\mu$
and the pion pole contribution $M_\pi$ as
\begin{equation}
	M^i_\mu = (-i)(-if_\pi q_\mu) \left ( \frac{i}{q^2 - m^2_\pi +
	i\epsilon} \right ) M^i_\pi(q) + \overline{M}^i_\mu \;. \label{eq:Mimu}
\end{equation}
To establish this (second) $S$-matrix form of PCAC, let i) $m_{\pi}^2
\approx 0$ in the pion propagator,  ii) take the divergence of both
sides of  (\ref{eq:Mimu}) and iii) use the Nambu version in the form
of  $q^\mu M^i_\mu \approx 0$ to arrive at
\begin{equation}
	i f_\pi M^i_\pi(q) = q^\mu \overline{M}^i_\mu
	\label{eq:smatpcac}
\end{equation}
Relation (\ref{eq:smatpcac}) can also be derived 
(~\cite{Scadronrev},pp 221) for 
$ m^2_\pi \neq 0$      with the aid of the
field theoretic statement $\partial A = f_\pi m^2_\pi \phi^i_\pi(x)$
where $\langle 0|\phi^i_\pi|\pi^j\rangle = \delta^{ij}$ and
$\phi^i_\pi(x)$ is {\em some} pseudoscalar field operator with the
quantum numbers of the pion.  It can be shown~\cite{Scadronrev} that equation 
(\ref{eq:smatpcac}) holds for either $m^2_\pi \rightarrow 0$ or 
$q^2 \rightarrow 0$, provided that the pion pole is first removed from 
$q^\mu M^i_\mu$.

The third and most useful form of PCAC (for our study of $\pi N$
scattering)  is  obtained from the soft-pion
limit ($q \rightarrow 0$) of (\ref{eq:smatpcac}) rewritten as 
$M^i_\pi(q) = {\textstyle{\frac{-i}{f_\pi}}}q^\mu \overline{M}^i_\mu$.
The right-hand side of this relation has contribution 
$\overline{M}^i_\mu \sim {\cal O}(1)$ which vanish as $q^\mu \rightarrow
0$.  However, the ${\cal O} (1/q)$ poles from ``tagging on" the
axial-vector to external nucleon lines will not vanish, giving the 
{\em soft pion theorem}:
\begin{equation}
	M^i_\pi(q)  \stackrel{q \rightarrow 0}{\longrightarrow}\;
	= \frac{-i}{f_\pi} q^\mu \overline{M}^i_\mu(\rm {poles})
	 + {\cal O}(q)\;,
\end{equation}
and the soft pion version of PCAC:  after removal of the pion poles and
${\cal O}(1/q)$ poles from the axial-vector amplitude the (truly)
background amplitude is a smoothly varying function of $q^2$ such that 
\begin{equation}
	q^\mu \overline{M}^i_\mu(\rm{non-pole}) \approx 0\;,  \nonumber
\end{equation}
and
\begin{equation}
	M^i_\pi(q^2)  \approx M^i_\pi(0)\;.   \label{eq:softpcac}
\end{equation}
This soft pion version of PCAC (\ref{eq:softpcac}) is now a statement
about {\em pion } amplitudes and can be used as such.  It is a sharp 
statement, comparable to other characterizations of PCAC, such as ``What
is special is that the pion mass is small, compared to the
characteristic masses of strong interaction physics; thus extrapolation
over a distance of $m^2_\pi$ introduces only small
errors"~\cite{Coleman}, pp 43.

\subsection{The Goldberger-Treiman Relation}
The Goldberger-Treiman (GT) relation between strong and weak
interaction parameters was displayed already in 1958~\cite{GT} and
explained by Nambu a short time later~\cite{Nambu}. Here we show that
the GT relation can be regarded as a single soft pion prediction of PCAC and pion pole
dominance of axial-vector, hadronic transitions.  First let us notice that the divergence of (\ref{eq:isoax})
coupled with the (Nambu) PCAC statement that  $\langle
N_{p'}|\partial A^i_\mu(x)|N_p\rangle \approx 0$ implies that the axial form
factors obey
\begin{equation}
	2 m g_A(q^2) + q^2 h_A(q^2) \approx 0 \;.  \label{eq:dum}
\end{equation}
where we have used $\gamma^\mu q_\mu \gamma_5 \rightarrow  2 m \gamma_5$,
when sandwiched between the spinors of on-mass-shell nucleons .  To go
farther, we dominate the axial-vector matrix element with the pion pole 
exactly as in Fig. 4, but this time we
have an effective $\pi NN$ coupling 
${\cal H}_{\pi NN} = g_{\pi NN}\bar{N}\vec{\tau}\cdot{\pi}\gamma_5 N$
which gives an explicit form to the pion pole $M^i_\pi(q)$ of 
(\ref{eq:Mimu}). Carrying this out we find
\begin{equation}
\langle N_{p'}|A^i_\mu(x)|N_p\rangle \approx 
	g_{\pi NN}\bar{u}_{p'}\tau^i\gamma_5
	u_p \frac{i}{q^2 - m^2_\pi +i\epsilon}(-i)(-if_\pi q_\mu)\;.
\end{equation}
Neglecting $m^2_\pi$ and comparing with (\ref{eq:isoax}) shows that it is
 the form factor  $h_A(q^2) i q_\mu\gamma_5$  which has the pion pole:
\begin{equation}
 	h_A(q^2) \approx -\frac{2 f_\pi g_{\pi NN}}{q^2}. \label{eq:app}
\end{equation}
Now let $q^2 \rightarrow 0$ to suppress the non-pion-pole terms, and the
$\approx$ in (\ref{eq:app}) becomes an equality, 
 turning (\ref{eq:dum}) into the exact  relation
\begin{equation}
   2m g_A(q^2=0) - 2 f_\pi g_{\pi NN}(q^2=0) = 0\;,
\end{equation}
which takes the familiar  Goldberger-Treiman form 
\begin{displaymath} 
 m g_A(0) = f_\pi g_{\pi NN}\;,
\end{displaymath}
our first soft pion prediction.

 To test the GT relation
   empirically in the chirally
broken real world, convert it to a Goldberger-Treiman {\em discrepancy}
\begin{equation}  
   \Delta = 1 - \frac{m_{N}g_{A}(0)}{f_{\pi}g}\;.
    \label{eq:gtdis}
 \end{equation}
The experimental values are~\cite{PDG}  
\begin{displaymath}  
m_{N} = \frac{1}{2}(m_{p} + m_{n}) = 938.91897 \pm 0.00028  \ \rm MeV,
 \end{displaymath} 
and~\cite{Holstein}
\begin{displaymath}  
  f_{\pi} = 92.6 \pm 0.2 \ \rm MeV,  
 \end{displaymath} 
We use the current best value of $g_{A}(t=0)$ as determined by two
consortia at the Institute for Nuclear Theory~\cite{Schiavilla98,RMP}.
 They find, by averaging modern results for the neutron 
lifetime and decay asymmetries,
\begin{displaymath}  
  g_{A}(0) = 1.2654 \pm 0.0042.  
 \end{displaymath} 
The least well known, and somewhat controversial, strong interaction
parameter is

\begin{eqnarray*}  
  g_{\pi NN} (q^2 = m^2_\pi) & \approx & 13.12 \;\;\;\cite{VPI} \\ 
  g_{\pi NN} (q^2 = m^2_\pi) & \approx & 13.02 \;\;\; \cite{Nijmegen} 
 \end{eqnarray*} 
down about $2\%$ from the pre-meson factory value of $g_{\pi NN} \approx
13.40$~\cite{ffrefs,Loiseau}.
The GT {\em discrepancy} then becomes
\begin{eqnarray}  
   \Delta & \approx & 0.023 \;\;\;\cite{VPI} \\ 
   \Delta & \approx & 0.015 \;\;\;\cite{Nijmegen}
    \label{eq:gtdisnum}
 \end{eqnarray}
or a discrepancy of only $2\%$!  This numerical fact is an, better than
usual, example of the soft pion form of PCAC; 
$M^i_\pi(q^2)  \approx M^i_\pi(0)$.

The Goldberger-Treiman relation is exact in the chiral limit $m^2_\pi
\rightarrow 0$ ($\partial A = 0$).  In our derivation we neglected 
 $m^2_\pi$ and then took the limit $q^2 \rightarrow 0$. Both limits are
 necessary to make the relationship exact.  This distinction becomes
 important as one attempts to use the $q^2 \rightarrow 0$ limit to guide
 the low $q^2$ variation of the $\pi NN$ vertex function for an off-mass
 shell pion in models of the $NN$ and $NNN$ forces~\cite{ffrefs}.  The
 value of $\Delta$ indicates a $2\%$ decrease in the coupling from the
 on-shell coupling $q^2 = m^2_\pi$ to $q^2 = 0$.  One should
 parameterize the $\pi NN$ ``form factor" to have this ``GT slope" which
 reflects chiral symmetry breaking.  The usual $\pi NN$ form factor
 of the Tucson-Melbourne $NNN$ force~\cite{TM79} has been parameterized
 to have about a $3\%$ GT slope.  That is, if $F_{\pi NN} (q^2) =
 (\Lambda^2 - m^2_\pi)/(\Lambda^2 - q^2)$ then $\Lambda \approx 800$
 MeV.
 
\subsection{The Adler Consistency Condition} 

Another soft pion result which is independent of current algebra
follows from an Adler-Dothan version of the soft pion theorem.  We
start with $M^i_\pi(q) = {\textstyle{\frac{-i}{f_\pi}}}q^\mu
\overline{M}^i_\mu$ for the general hadronic amplitude $A \rightarrow B
+ \pi^i$, and examine the origin of the ${\cal O}(q^{-1})$ 
nucleon poles which survive in the limit $q \rightarrow 0$. Then
 only the axial vector (designated by $\star$ ``tagging"
onto external nucleon lines of the general hadronic ingoing line  $A$
and outgoing line $B$ in the diagram below) will generate nucleon
propagators ${\cal O}(q^{-1})$.  The $\star$
represents a $g_A(q^2)$-type coupling of the axial vector, since 
$h_A(q^2)$ is already included in $M^i_\pi(q^2)$ from the pion pole in 
$h_A(q^2)$, see (\ref{eq:app}).

\begin{center}
\begin{picture}(350,100)(0,0)
\ArrowLine(64.142,64.142)(100,100)
\Text(72,15)[]{$A$}
\ArrowLine(100,0)(64.142,35.858)
\Text(75,90)[]{$B$}
\CCirc(50,50){20}{Black}{White} 
\Text(90,90)[]{\Large$\star$}
\Text(110,50)[]{$+$}
\SetOffset(100,0)
\ArrowLine(64.142,64.142)(100,100)
\Text(72,15)[]{$A$}
\ArrowLine(100,0)(64.142,35.858)
\Text(75,90)[]{$B$}
\CCirc(50,50){20}{Black}{White} 
\Text(90,10)[]{\Large$\star$}
\end{picture}
\end{center}
 
Now we apply the GT relation $m g_A(0) = f_\pi g_{\pi NN}$ to identify
the nucleon pole parts of the axial background (i.e.,pion-pole removed)
amplitude $q^\mu \overline{M}^i_\mu$ with the pseudoscalar $\pi N$
interaction, so that the diagrammatic representation of the background
becomes:
\begin{center}
\begin{picture}(350,100)(0,0)
\ArrowLine(64.142,64.142)(100,100)
\Text(72,15)[]{$A$}
\ArrowLine(100,0)(64.142,35.858)
\Text(75,90)[]{$B$}
\CCirc(50,50){20}{Black}{White} 
\DashLine(90,90)(93,77){3}
\Text(110,50)[]{$+$}
\Text(210,50)[]{$+$}
\Text(270,50)[]{$( M_0\gamma_5 \tau^i + \gamma_5 \tau^i M_0)$}
\SetOffset(100,0)
\ArrowLine(64.142,64.142)(100,100)
\Text(72,15)[]{$A$}
\ArrowLine(100,0)(64.142,35.858)
\Text(75,90)[]{$B$}
\CCirc(50,50){20}{Black}{White} 
\DashLine(90,10)(93,23){3}

\end{picture}
\end{center}

The hadronic amplitude $A \rightarrow B$ labeled $M_0$ contains  nucleons
but has had, as we have seen, the soft pion removed and the axial vector
removed. Now let $q^\mu \rightarrow 0$  in 
 $M^i_\pi(q) = {\textstyle{\frac{-i}{f_\pi}}}q^\mu \overline{M}^i_\mu$,
 to suppress all further background parts  in $\overline{M}^i_\mu$ of 
${\cal O}(q^0)$.  Only the nucleon poles ${\cal O}(q^{-1})$ with pseudoscalar 
pion-nucleon coupling are left and we have the soft pion theorem proved
by Adler and Dothan~\cite{AD}: 
\begin{equation} 
 M^i_\pi(q) \stackrel{q \rightarrow 0}{\longrightarrow}\; 
  M^i_{\rm ps N poles}(q) + \overline{M}^i_\pi(q \rightarrow 0)
\end{equation} 
where 
 \begin{equation}
 \overline{M}^i_\pi(q \rightarrow 0) = \frac{g_{\pi NN}}{2m}
    ( M_0\gamma_5 \tau^i + \gamma_5 \tau^i M_0)\;. \label{eq:Mpibar}
\end{equation}  
This version of the soft pion theorem, valid for either an incoming or
an outgoing soft pion, is the analog of the soft photon theorem of 
Low~\cite{Low}.  It allows us to turn  the Adler
zero~\cite{Adler65}, $M^i_\pi(q) \stackrel{q \rightarrow
0}{\longrightarrow}\; 0$ provided that $\overline{M}^i_\mu$ in
(\ref{eq:smatpcac}) has no poles, into the Adler PCAC consistency condition
for $\pi N$ scattering.

To begin this demonstration, let us display explicitly the $s$-channel
and $u$-channel {\em nucleon} poles in $\pi^j(q) + N(p) \rightarrow 
\pi^i(q') + N(p')$:

\begin{figure}[htbp]
\unitlength1.cm
\begin{picture}(6,3.5)(2.5,15)
\includegraphics{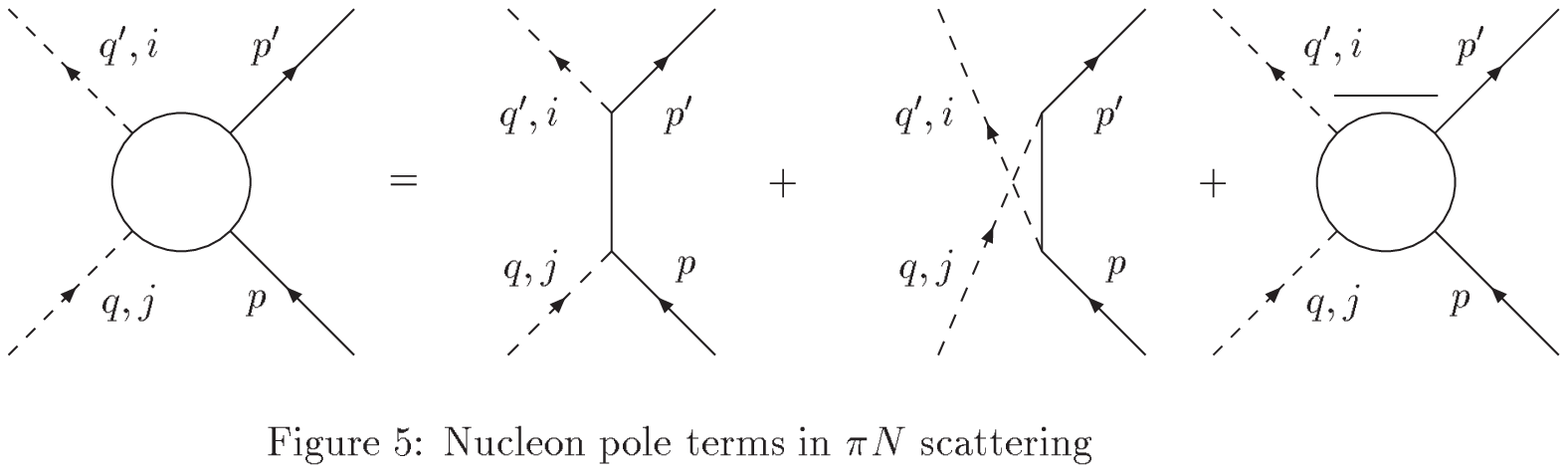}
\end{picture}
\end{figure}

Both nucleon poles are present and are added together by the Feynman rules
because the crossed pions are bosons.  Now let the final pion become
soft ($q' \rightarrow 0$) and the other three particle be on-mass-shell.
As we have separated out the pseudoscalar nucleon poles, we can apply
(\ref{eq:Mpibar}) with $M_0 = -g_{\pi NN}\tau^j\gamma_5$  from
$N\rightarrow N + \pi$.  Then
\begin{eqnarray}
\overline{M}^{ij}_\pi(q \rightarrow 0) &=& -\frac{g_{\pi NN}}{2m}
    ( g_{\pi NN}\tau^j\gamma_5\gamma_5 \tau^i + \gamma_5 \tau^i 
 g_{\pi NN}\tau^j\gamma_5)\\
        &=& + \frac{g^2_{\pi NN}}{m} \delta^{ij}  \label{eq:acc}
\end{eqnarray}
where we have used $\{\tau^i, \tau^j\} = 2 \delta^{ij} $ and remind the
reader that $\gamma^2_5 = -1$ in this (Schweber) convention.

Now we restate the Adler
consistency condition (\ref{eq:acc}) as a condition on the invariant
amplitude $F^+ = A^+ + \nu B^+$  (since the isospin condition is
$t$-channel even).  The kinematic variables for $q^2 = m^2_\pi = \mu^2$ 
and $q' \rightarrow 0$ are $t = (q^2 - q'^2)= \mu^2$ and $\nu =
 0$ because $s=u=m^2$.  Then the Adler consistency condition
becomes
\begin{equation}
F^+(\nu=0,t=\mu^2;q^2=\mu^2,q'^2=0) =
A^+(\nu=0,t=\mu^2;q^2=\mu^2,q'^2=0) = \frac{g^2}{m}\;\;,
\label{eq:adlerf}
\end{equation}
and we note the  pseudoscalar nucleon poles do {\em not} contribute to  
$A^{(\pm)}$ but only to $B^{(\pm)}$.  Specifically, 
\begin{equation}
A^{(\pm)}_{ps} = 0 \hspace{30pt} B^+_{ps} = 
\frac{g^2}{m} \frac{\nu}{\nu_B^2 - \nu^2} \hspace{30pt}
B^-_{ps} = \frac{g^2}{m} \frac{\nu_B}{\nu_B^2 - \nu^2}
\end{equation}
To make contact with $\pi N$ data analyses obtained from dispersion
relations, it is natural to evaluate pseudoscalar nucleon poles, not as
field theory Feynman diagrams but in the sense of dispersion theory so
that the residue in $\nu^2$ of $F^+_P$ is evaluated at the value of
$\nu$ at the $s$-channel nucleon pole, $2m(\nu_B - \nu) = (m^2 -s)$ or
$\nu_B = -q\cdot q' /2m$: 
\begin{equation}
F^+_P = \frac{g^2}{m} \frac{\nu_B^2}{\nu_B^2 - \nu^2}\hspace{50pt}
F^-_P = \frac{g^2}{m} \frac{\nu \nu_B}{\nu_B^2 - \nu^2}\;\;,\label{eq:nucpole}
\end{equation}
(see Ref. \cite{Scadronbook}, pp 340-343).  The difference between the two
prescriptions lies only in $F^+$:
\begin{equation}
	F^+_P(\nu,t) = F^+_{ps}(\nu,t) + \frac{g^2}{m}\;.
\end{equation}
Now  restate the Adler consistency condition in the
form of a condition on the {\em background} $\pi N$ amplitude defined as
\begin{equation}
	F^{(\pm)} = F^{(\pm)}_P + \bar{F}^{(\pm)}\;,  \label{eq:Fbar}
\end{equation}
so that 
\begin{equation}
	\bar{F}^+ = \bar{F}^+_{ps} - \frac{g^2}{m}\;.
\end{equation}
In the single soft (Adler) limit (\ref{eq:adlerf}) becomes
\begin{equation}
	F^+_P \rightarrow 0\;, \hspace{50pt} 
	\bar{F}^+ \rightarrow \frac{g^2}{m} - \frac{g^2}{m} =0.
	\label{eq:Adlerzero}
\end{equation}
The knowledgeable reader may have noticed that for $F^+$ 
the dispersion-theoretic
nucleon pole with pseudoscalar coupling is the same as the field
theoretic nucleon pole of $F^+$ with pseudovector coupling, so one
can think, if one wishes, think of the
background $\bar{F}^+ $ as the full amplitude minus the
pseudovector poles.  It is often said that the Adler PCAC consistency
condition of chiral symmetry forces the use of pseudovector coupling,
but it is obvious from the above that this soft pion theorem makes no
such demand.
In the future, the phrase ``nucleon poles" refer to dispersion theory
poles with pseudoscalar coupling.

Invoking PCAC in the form of (\ref{eq:softpcac}), we can expect that
putting the final pion back on-mass-shell (and holding fixed $t$ and
$\nu$) should not change the Adler
consistency condition much:
\begin{equation}
\bar{F}^+(0,\mu^2) \equiv \bar{F}^+(\nu=0, t=\mu^2; q^2=\mu^2, q'^2=\mu^2) 
\approx 0\;.
	\label{eq:adlerlet}
\end{equation}
This ``Adler Low Energy Theorem" (LET) point is in the 
subthreshold crescent region of the
Mandelstam plane (see Fig. 2) and the value of $\bar{F}^+$ can be reliably
determined from $\pi N$ scattering data with the aid of dispersion
relations.  It is $\bar{F}^+ \approx -0.03 \mu^{-1}$~\cite{Pavan} or 
$\bar{F}^+ \approx -0.08 \mu^{-1}$~\cite{Kaufmann}, extrapolations obtained
 from the most recent
phase shift analysis called SM98~\cite{SM98}.  As the amplitude, unlike
the Goldberger-Treiman discrepancy, has dimensions we must compare this
result to the overall scale  $-1.3\mu^{-1} \leq \bar{F}^+(\nu,t) \leq 6
\mu^{-1}$ within the subthreshold crescent. Then we see that this PCAC low energy
theorem (\ref{eq:adlerlet}) is also rather
impressively confirmed by the data~\cite{olddata}.  Indeed one can go a
step further and notice that this background amplitude has a zero in
the subthreshold crescent  which, beginning at the Adler LET,  passes
very near the threshold point  ($\nu = \mu, t=0$)~\cite{Kaufmann}.  The
nucleon pole contribution is quite small ($\approx  -0.13 \mu^{-1}$) at
threshold, so the overall {\em unbarred} isoscalar scattering length
$a_0 \approx F^+(\mu,0)/4\pi\approx 0.01\mu^{-1} \approx 0$ 
is a threshold consequence of
the PCAC Adler consistency condition and has nothing to do with current
algebra.

\subsection{Current Algebra and $\pi N$ Scattering}

The current algebra representation of low-energy $\pi N$ scattering not
only utilises on-mass-shell ($q^2 = q'^2 = \mu^2$) axial Ward-Takahashi
identities (analogous to the conditions gauge invariance imposes on
photon-target scattering, but incorporating a current algebra
commutation relation) but also make a specific prediction for the
amplitude $\pi^j(q) + N(p) \rightarrow  \pi^i(q') + N(p')$ when both
pions are soft~\cite{Weinberg66}.  This latter prediction can be
compared to the data by invoking  PCAC in the form of
(\ref{eq:softpcac}) to bring each pion back to the mass-shell. But first
we must establish this current algebra
representation~\cite{Adler65,Scadronrev}.

\begin{center}
\vspace{10pt}
\begin{picture}(100,100)(0,0)
\Photon(0,0)(35.858,35.858){3}{3}
\Text(35,15)[]{$A^j_\nu(q)$}
\ArrowLine(64.142,64.142)(100,100)
\Text(72,15)[]{$p$}
\Photon(35.858,64.142)(0,100){3}{3}
\Text(35,90)[]{$A^i_\mu(q')$}
\ArrowLine(100,0)(64.142,35.858)
\Text(75,90)[]{$p'$}
\CCirc(50,50){20}{Black}{White} 
\end{picture}
\\ \mbox{} \\ Figure 6: Compton-like two-current scattering diagram.
\end{center}

Begin with $SU(2)$ axial currents ``scattering" off target nucleons and
write the covariant amplitude (to be sandwiched between on-shell 
nucleon spinors) as
\begin{equation} 
	M^{ij}_{\mu\nu} = i \int d^4x e^{iq'\cdot x} 
	T[A^i_\mu(x),A^j_\mu(0)]\theta(x_0)\;,  \label{eq:M}
\end{equation}
where $\Delta = q-q' = p'-p$ and the momentum transfer is $t=\Delta^2$.
Contract (\ref{eq:M}) with $q'$ (i.e. take a divergence in coordinate
space),  integrate the right hand side by parts, and drop the surface
term at infinity. Using the identity 
$\partial^\mu T(A_\mu(x)\ldots) = T(\partial A\ldots) +
\delta(x_0)A_0\ldots$,  we find
\begin{equation}
	q'^\mu M^{ij}_{\mu\nu}	= i \int d^4x e^{iq'\cdot x}
	T[\partial^\mu A^i_\mu(x),A^j_\mu(0)]\theta(x_0)
	- i \epsilon^{ijk} \Gamma^k_\nu(\Delta)\;,  \label{eq:qM}
\end{equation}
where we have used the Equal Time Commutation relationship
(\ref{eq:curralg})
to bring in the three-point vertex function $\Gamma^k_\nu(\Delta)$ which
depends only on the momentum transfer $\Delta = (p' - p)$.

\begin{center}
\begin{picture}(100,100)(0,0)

\ArrowLine(64.142,64.142)(100,100)
\Text(72,15)[]{$p$}
\ArrowLine(100,0)(64.142,35.858)
\Text(75,90)[]{$p'$}
\CCirc(50,50){20}{Black}{White} 
\Photon(-10,50)(30,50){3}{3}
\Text(10,55)[b]{$J^k$}
\PText(1,51)(45)[]{>}
\Text(20,10)[]{$\Gamma^k_\nu(\Delta)$}

\end{picture}
\\  \mbox{} \\ Figure 7: The three-point vertex $M$ function for 
the isovector-vector current.
\end{center}

Before going on, let us pause to examine the extension to currents of
the charge algebra of (\ref{eq:QQ}) and (\ref{eq:QQ5}):
\begin{eqnarray}
	[Q^i,J^j_{\nu}(x)]_{ETC} &=& i f^{ijk}J^k_\nu(x) \nonumber \\ 
	\mbox{}     [Q^i,A^j(x)]_{ETC}  &=&  i f^{ijk}A_\nu^k(x) \nonumber \\ 
 	\mbox{}     [Q^i_5,J^j(x)]_{ETC}  &=&  i f^{ijk}A_\nu^k(x) \nonumber \\ 
	\mbox{}     [Q^i_5,A^j(x)]_{ETC}  &=&  i f^{ijk}J^k_\nu(x)\;. 
		\label{eq:curralg}
\end{eqnarray}
The axial charge is, of course, defined as $Q^i_5 = \int d^3x
A_0^i(t,\vec{x})$, by analogy to $Q^i = \int d^3x J_0^i(t,\vec{x})$,
and the $SU(3)$ structure constants $f^{ijk}$ reduce to
$\epsilon^{ijk}$ for $i = 1,2,3$ of $SU(2)$ pion-nucleon scattering.  Then one
can recover the charge algebra (\ref{eq:QQ}) and (\ref{eq:QQ5}) from
the current algebra relations (\ref{eq:curralg}) by setting $\nu = 0$
and integrating over all space.  Notice that if the currents in
(\ref{eq:M}) were the {\em conserved} isovector-vector current
$J_{nu}^j(k)$ and $J_{mu}^i(k')$ then (\ref{eq:qM}) would become
simply  $k'^\mu M^{ij}_{\mu\nu} = - i \epsilon^{ijk}
\Gamma^k_\nu(\Delta)$. The latter  is the isovector-vector Ward-Takahashi
identity for virtual isovector photons which replaces the gauge 
invariance equation for real photons.

Now we go on, by contracting (\ref{eq:qM}) by $q^\nu$ and converting the
$x$ dependence of the currents from $\partial A^i$ to $A^j_\nu$ so that
we can integrate by parts once again.  The result is the ``double"
Ward-Takahashi identity
\begin{eqnarray}
	\lefteqn{q'^\mu M^{ij}_{\mu\nu}q^\nu	= }  \\
	& & \int d^4x e^{-iq\cdot x}
	T[\partial^\mu A^i_\mu(0),A^j_\mu(x)]\theta(x_0) 
	 - i \epsilon^{ijk} \Gamma^k_\nu(\Delta)q^\nu + 
	[i \partial^\mu A^i_\mu(0),Q^j_5]_{ETC} \nonumber
			\label{eq:qMq}
\end{eqnarray}
We can symmetrise the current algebra term 
$i \epsilon^{ijk} \Gamma^k_\nu(\Delta)q^\nu = 
i \epsilon^{ijk} \Gamma^k_\nu(\Delta)Q^\nu$ by utilising 
$\Gamma^k_\nu(\Delta)\Delta^\nu = 0$ and the definition $Q = {\textstyle
{\frac{1}{2}}} (q + q')$.   This term can be identified with the
measured electromagnetic form factors of nucleons~\cite{emff}:
\begin{equation}
\Gamma^k_\nu(\Delta)Q^\nu = {\textstyle{\frac{\tau}{2}}}\left[
	F^V_1(t) \gamma_\nu Q^\nu - {\textstyle{\frac{1}{4m}}}
        F^V_2(t)[\gamma^\nu q'_\nu, \gamma^\nu q_\nu]\right]\;,
\end{equation}
where we have put back in the suppressed nucleon spinors and used the
defining equation (\ref{eq:isovec}).  The second commutator term on the
RHS of (\ref{eq:qMq}), reinstating the nucleon spinors, is the
pion-nucleon ``sigma" term
\begin{equation}
	\langle N_{p'}|[i \partial^\mu A^i_\mu(0),Q^j_5]_{ETC}|N_p\rangle
	= \delta^{ij} \sigma_N(t)\bar{N}_{p'}N_p\;, \label{eq:sigterm}
\end{equation}
which like the current algebra term can be only a function of $t$.  The
amount of the $t$ dependence of the sigma term cannot be determined by
theory,  but like the $t$ dependence of the the current algebra term is
obtained by comparing with measurement.  The sigma term is isospin
symmetric, as can   can be established by reversing the order of
momentum contractions of $M^{ij}_{\mu\nu}$.  The sigma term is a
measure of chiral symmetry breaking since it is proportional to the
non-conserved current $\partial A \neq 0$.

Having discussed the two $t$ dependent terms on the RHS of
(\ref{eq:qMq}), we now relate the LHS to pion-nucleon scattering by
dominating the LHS by pion poles according to Figure 8.

\begin{figure}[htbp]
\unitlength1.cm
\begin{picture}(6,3.5)(2.5,16)
\includegraphics{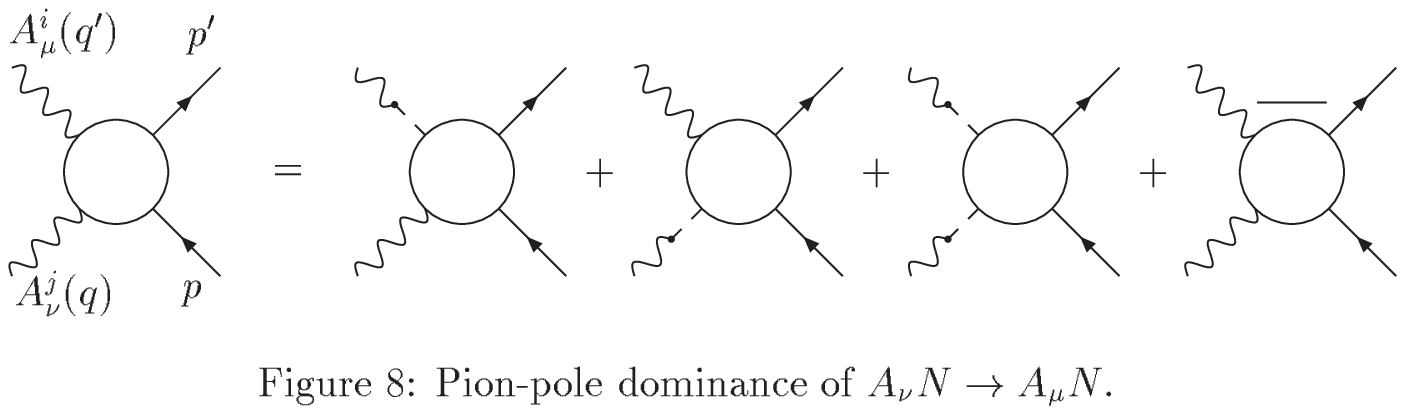}
\end{picture}
\end{figure}

The $S$-matrix version of PCAC $i f_\pi M^i_\pi(q) = q^\mu
\overline{M}^i_\mu$  holds no matter if we let $\mu ^2 \rightarrow 0$
or $q^2 \rightarrow 0$, or keep both non-zero.  So we can 
 take the double divergence of Fig. 8 as in (\ref{eq:qMq}),
first setting $\partial A=0$, which
gets rid of the integral on the RHS of (\ref{eq:qMq}), and equate the
result to the RHS of (\ref{eq:qMq}), giving
\begin{equation}
	- f_\pi ^2 M^{ij}_{\pi \pi}  + q'^\mu \bar{M}^{ij}_{\mu\nu}q^\nu 
	= -i \epsilon^{ijk}
{\textstyle{\frac{\tau ^k}{2}}}\left[
	F^V_1(t) \gamma_\nu Q^\nu - {\textstyle{\frac{1}{4m}}}
        F^V_2(t)[\gamma^\nu q'_\nu, \gamma^\nu q_\nu]\right]
  + \delta^{ij} \sigma_N(t)\;.   \label{eq:qMq2}
\end{equation}
The minus sign of $M_{\pi \pi}$ is the reverse of the sign associated
with the third (the one we really want) diagram on the RHS of the figure
because of the first two diagrams.

This relationship between  $\pi N$ scattering and a double divergence
has been obtained from the covariant amplitude (\ref{eq:M}) with the aid
of pion-pole dominance and the S-Matrix version of PCAC
(\ref{eq:smatpcac}). 
In order to reproduce two noteworthy current algebra/PCAC
theorems, the Weinberg double soft-pion theorem~\cite{Weinberg66} and
the Adler-Weisberger double soft-pion relation~\cite{AW} we now use
the  soft pion theorem (\ref{eq:softpcac}).  The latter can be applied
only if {\em all} poles are removed from {\em both} amplitudes: 
$M^{ij}_{\pi \pi}$ and the non-pion pole axial-Compton amplitude 
$\bar{M}^{ij}_{\mu\nu}$. 
The nucleon poles are removed from 
$M^{ij}_{\pi \pi}$ as in Fig.  5 and in an analogous figure for nucleon
poles in $\bar{M}^{ij}_{\mu\nu}$, with the important distinction that
the nucleon-axial coupling is not pseudoscalar 
$g_{\pi NN}\gamma_5\equiv g \gamma_5$ but instead is defined by
({\ref{eq:isoax}). This distinction 

\begin{equation}
	-  M^N_{\pi \pi} + f_\pi^{-2} q'^\mu \bar{M}^N_{\mu\nu}q^\nu 
	= \delta^{ij}\frac{g^2}{m} + i \epsilon^{ijk}\tau^k \frac{g^2
	\nu}{2m^2}
\end{equation}
introduces the Adler contact term back into the isospin-even $\pi N$
amplitude.
With the removal of the
nucleon poles the resulting 
$q'^\mu \bar{M}'^{ij}_{\mu\nu}q^\nu$ vanishes as 
$q \rightarrow 0\;,q' \rightarrow 0$. 
The upshot is the generic double soft-pion result:
\begin{equation}
	M_{\pi \pi}(q,q') \approx M_{\pi \pi}^{ps N}(q,q')  +
	\bar{M}_{\pi \pi}(q \rightarrow 0\;,q' \rightarrow 0)\;,
\end{equation}
which has the isospin decomposition ({\ref{eq:isospin})
\begin{equation}
	\bar{M}_{\pi \pi}^+(q \rightarrow 0\;,q' \rightarrow 0) = 
		\frac{g^2}{m} - \frac{\sigma_N(0)}{f_\pi ^2}\;,
		\label{eq:Wpt}
\end{equation}
derived by Weinberg~\cite{Weinberg66}, and
\begin{equation}
	\nu ^{-1} \bar{M}_{\pi \pi}^-(q \rightarrow 0\;,q' \rightarrow 0) = 
\frac{1}{f_\pi ^2} - \frac{g^2}{2m^2} = \frac{1}{f_\pi ^2}(1 -
g^2_A)\;,
	\label{eq:AWpt}
\end{equation}
the Adler-Weisberger relation~\cite{AW}.  To obtain the crossing
symmetric relation (\ref{eq:AWpt}), we divide  $F^V_1(t) \gamma_\nu
Q^\nu$ of (\ref{eq:qMq2}) by $\nu = (p + p')\cdot(q + q')/4m$, before
taking the $q \rightarrow 0,q' \rightarrow 0$ limit, to get 
${\textstyle{\frac{1}{2}}} F^V_1(t)\rightarrow
{\textstyle{\frac{1}{2}}}$ because $F^V_1(0)= 1$.  We then use the GT
relation to obtain the  far RHS of 
(\ref{eq:AWpt}).

\subsection{On-pion-mass-shell current algebra Ward-Takahashi identities}

We have derived the Ward identities in the chiral limit such that
 $q'^\mu \bar{M}'^{ij}_{\mu\nu}q^\nu$ vanishes
as  $q \rightarrow 0,q' \rightarrow 0$, where $\bar{M}'^{ij}_{\mu\nu}$
is the background axial-Compton amplitude with pion and nucleon poles
removed.
A derivation similar to the above, but keeping both pions on-mass-shell
 at all stages, yields  on-shell Ward identities which impose 
 current algebra constraints on pion-nucleon scattering~\cite{BPP}. 
 These identities are most conveniently expressed by writing the (not
 necessarily zero) double divergence in the same manner as the $M$
 function for $\pi N$ scattering (\ref{eq:FandB})
\begin{eqnarray}
 	M_{\pi \pi}^{(\pm)} &=& F^{\pm}(\nu,t)
- {\textstyle{\frac{1}{4m}}}[\rlap/q,\rlap/q']B^{\pm}(\nu,t)
\nonumber \\
q'^\mu \bar{M}'^{\pm}_{\mu\nu}q^\nu &=& C^{\pm}(\nu,t) 
- {\textstyle{\frac{1}{4m}}}[\rlap/q,\rlap/q']D^{\pm}(\nu,t)
\end{eqnarray}
Now define the background $\pi N$  amplitude as $\bar{M}\equiv M -M_P$
where $M_P$ is the {\em dispersion-theoretic} pole of
(\ref{eq:nucpole}),
see Figs. (2) and (5), for pseudoscalar $\pi NN$ coupling.  The
on-pion-mass-shell Ward-Takahashi identities take the form 

\begin{eqnarray}
  \bar F^{+}(\nu ,t) &=&  \frac{\sigma_N(t)}{f_{\pi}^2}
      + C^{+}(\nu ,t)
 \label{eq:fplus}  \\
 \nu^{-1} \bar F^{-}(\nu ,t) &=& \frac{F^V_1(t)}{2f_{\pi}^2} - 
\frac{g^2}{2m^2} + \nu^{-1} C^{-}(\nu ,t)
 \label{eq:fminus}  \\
 \nu^{-1} \bar B^{+}(\nu ,t) &=& \nu^{-1} D^{+}(\nu ,t)
 \label{eq:bplus}  \\
 \bar B^{-}(\nu ,t) &=& \frac{F^V_1(t) + F^V_2(t)}{2f_{\pi}^2} - 
\frac{g^2}{2m^2} +  D^{-}(\nu ,t)\;,
 \label{eq:bminus}
\end{eqnarray}
where we notice  that removal of the dispersion theory pole removes the
contact term $g^2/m$ from (\ref{eq:fplus}). The on-shell analogue
({\ref{eq:fminus}) of the Adler-Weisberger double soft pion point
({\ref{eq:AWpt}) goes to ({\ref{eq:AWpt}) because $C^-(\nu,t)$
is {\em defined} as a double divergence in coordinate space which
vanishes as ($q\rightarrow 0, q'\rightarrow 0$).  Comparing with the
double-soft pion Weinberg limit (\ref{eq:Wpt}), which can be written as 
\begin{equation}
        \bar{F}^+ (q\rightarrow 0, q' \rightarrow 0) = 
	- \frac{\sigma_N(0)}{f_\pi ^2} \:,
\label{eq:ca}
\end{equation}
we note that the sign change is due to an analytic power series
expansion in $q$ and $q'$ (scaled to a typical hadron mass such as
$m$) which obeys the Adler zero ({\ref{eq:Adlerzero})~\cite{CD}.  In
fact,
Brown, Pardee, and Peccei~\cite{BPP} suggest the  
sigma-term structure $\sigma_N(t)[(q^2 + q'^2)/\mu^2 -1]$ which manifests
the sign change and the Adler zero.  They also confirm the on-shell
Cheng-Dashen (CD) low energy theorem of 1971 \cite{CD} :
\begin{equation}
        \bar{F}^+(0,2\mu^2)  = \frac{\sigma_N(2\mu^2)}{f_\pi ^2} + 
	{\cal O}(\mu^4)\;.   \label{eq:cd}
\end{equation}

With the on-shell Ward identities we are now in a position to test the
current algebra representation with the data of $\pi N$ scattering
extrapolated to the sub-threshold crescent of Fig. 2.  These tests can
establish the magnitude of the sigma term in (\ref{eq:fplus}), suggest
(with the aid of the PCAC hypothesis) the $t$ dependence of the sigma
term,  and confirm or deny the $t$ dependence of the current algebra
terms  (from (\ref{eq:curralg}))  in (\ref{eq:fminus}) and 
(\ref{eq:bminus}).  We turn to these tests in the next section.

\section{Tests of Current Algebra and Soft Pion Theorems} 

In this lecture we use contemporary analyses of  on-mass-shell $\pi N$
scattering to i) test the structure of current algebra in this context,
and to ii) examine the validity of the PCAC hypothesis (already
validated for the Adler LET (\ref{eq:adlerlet})) that the exact double
soft-pion theorems of Weinberg (\ref{eq:Wpt}) and  of Adler and
Weisberger (\ref{eq:AWpt})  should be evident in the $\pi N$ data.  The
former tests of current algebra were initiated by the data analysis of
the Karlsruhe group~\cite{HJS} which use fixed-$t$ dispersion relations
to extrapolate from the(ir) $s$-channel experimental phase shifts into
the subthreshold region around ($\nu=0,t=0$) (see Fig. 1).  In fact,
the experimental information  in this region, once the nucleon pole
contributions (see Fig. 2) have been removed, can be expressed in terms
of the expansion coefficients of the four {\em background} $\pi N$
invariant amplitudes about this point. The values of these  H\"{o}hler
expansion coefficients obtained from  data taken  in the 1970's, before
the meson factories were built, are summarized in the encyclopedic
Ref. \cite{hohlerbook}.  The two current algebra models~\cite{ST,OO}
of   $\pi N$ amplitudes, which have been adapted to the construction of
two-pion exchange three-nucleon forces [8-11], were calibrated against
these pre-meson factory H\"{o}hler coefficients.

Recently two of the twenty-eight subthreshold subthreshold coefficients
have been re-evaluated~\cite{Pavan} via fixed-$t$ dispersion relations
from the latest  partial wave analysis~\cite{SM98} of meson factory
data.   We will use the more comprehensive
determination~\cite{Kaufmann} of the amplitudes $\bar F^{+}(\nu ,t)$
and $\nu^{-1} \bar F^{-}(\nu ,t)$  in the subthreshold crescent to
fully carry out the tests i) and ii) of chiral symmetry.  This
determination also is from the  VPI phase-shift analysis
SM98~\cite{SM98}, but the analysis used interior dispersion relations
(IDR), pioneered by Hite, et al.~\cite{Hite}, and advocated by 
H\"{o}hler for the purpose of testing chiral symmetry~\cite{GH94}. 
Interior dispersion relations are evaluated along hyperbolas in the
Mandelstam plane which correspond to a fixed angle in the $s$-channel
laboratory  frame. For $t < 0$ the path of fixed lab angle lies
entirely within the  $s$-channel physical region and passes through the
$s$-channel threshold point.  (The IDR paths are similar to the lines
of fixed center of mass  three-momentum $q^2$ and $x = \cos
\theta_{cm}$ of Fig. 9, from Ref. ~\cite{Murphy} and illustrating a
construction of invariant amplitudes from a simple summation of
partial-wave amplitudes continued into the subthreshold crescent). With
the IDR paths, one can reliably extrapolate the invariant amplitudes to
any point in the subthreshold crescent and {\em in particular} evaluate
the amplitudes along the vertical axis ($\nu = 0, 0\leq t \leq 4\mu^2$)
to test the on-shell analogues of the soft pion theorems. 
\pagebreak

\setcounter{figure}{8}
\begin{figure}[htpb]
\unitlength1.cm
\begin{picture}(5,10)(-14,0.8)
\includegraphics{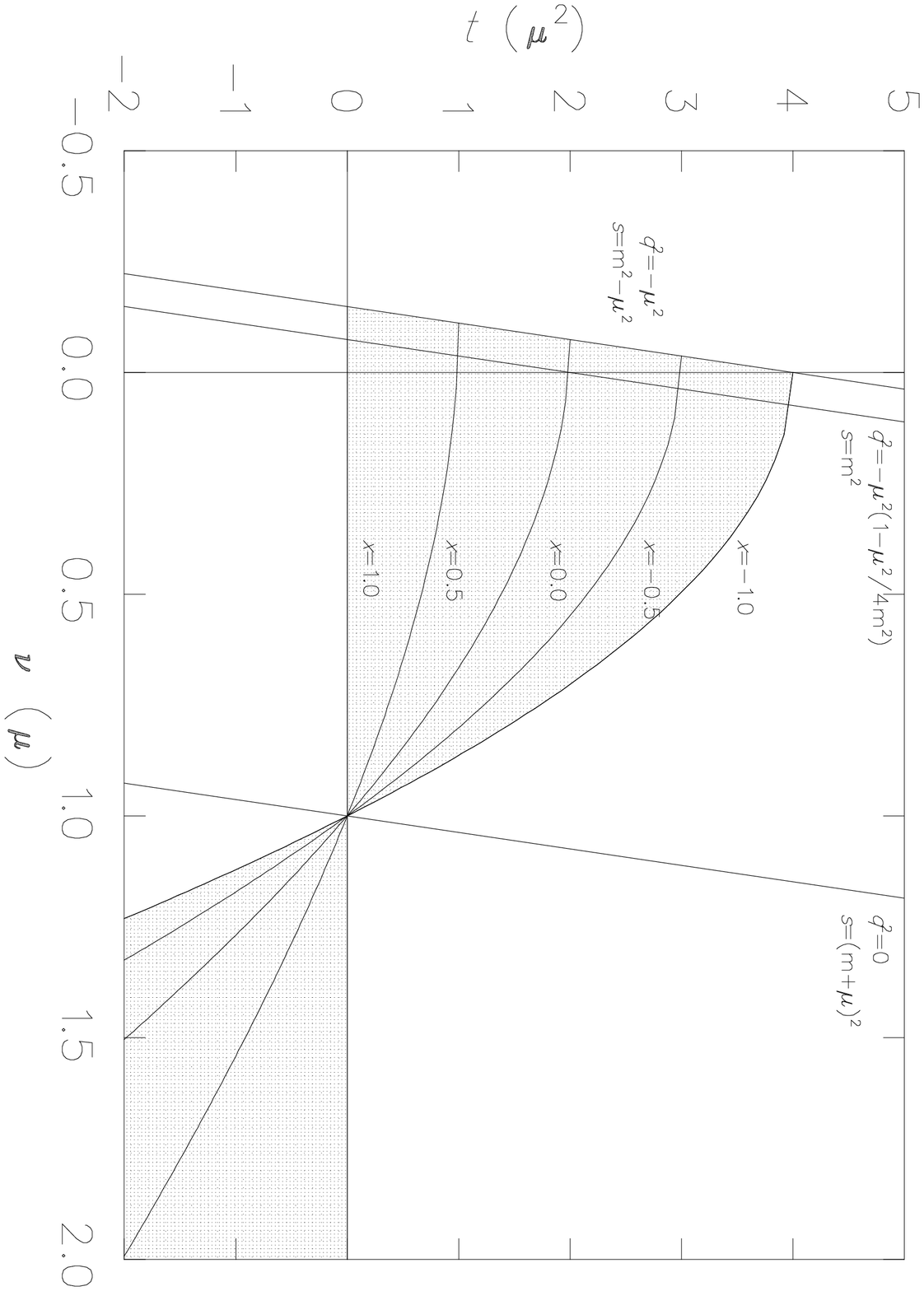}
\end{picture}
\caption{
  A portion of the (all four particles on-mass-shell) 
 Mandelstam ($\nu,t$) plane, which includes the $s$-channel physical
 region and the subthreshold crescent. }
\end{figure}

But first we must attempt to calibrate the IDR invariant amplitudes
from SM98 phase shifts against independent measurements.   The IDR
value  at the $s$-channel threshold point $(\nu = \mu, t=0$) the
 scattering lengths
\begin{eqnarray*}
  a^{(+)} & = & \frac{1}{4\pi(1 + \mu/m)}F^+(\mu,0) \approx -0.005\;
  \mu^{-1} \\
    a^{(-)} & = & \frac{1}{4\pi(1 + \mu/m)}F^-(\mu,0) \approx +0.087\; \mu^{-1} 
\end{eqnarray*}
in good agreement with the preliminary values  $a^{(+)} = .0016 \pm
.0013\; \mu^{-1}$ and $a^{(-)} = 0.0868\pm .0014\; \mu^{-1}$ from the 1s
level shifts and widths in pionic hydrogen and deuterium~\cite{atom}.
In addition, at the pseudothreshold point $\nu = 0$, $t=4\mu^2$ 
(a focus of the boundary hyperbola; see Fig. 1) the $I=0$ $\pi \pi$
scattering length, $a_{00}$, can be evaluated  with the
method of Ref.~\cite{IDRpipi}.  The IDR value $a_{00} \approx 0.20
\mu^{-1}$ again agrees well with the recent determination of 
$a_{00} = 0.204 \pm 0.014 \pm 0.008\; \mu^{-1}$ from the totally independent
reaction $\pi N \rightarrow \pi \pi N$~\cite{CHAOS}.  That the IDR give
reasonable values of $a_{00}$ and $a^{(+)}$ at opposite points on the boundary
of the subthreshold crescent adds confidence to the values in the
central region.  

Our first test of a combined current algebra-PCAC prediction
(\ref{eq:AWpt}) uses the IDR
determination of the isospin odd amplitude $\nu^{-1} \bar F^{-}(\nu ,t)$
at the point $(\nu = 0, t=0)$.  The exact current algebra
Adler-Weisberger result (\ref{eq:AWpt}) is 
\begin{equation}
	\nu^{-1} \bar F^{-}(q\rightarrow 0,q'\rightarrow 0) = 
	\nu^{-1} \bar F^{-}(0,0;0,0) = \frac{1}{f_\pi ^2}(1-g^2_A)\
	= -0.62 \; \mu^{-2}\;,  \label{eq:AWvalue}
\end{equation}
where we have used the values of section 3.4.  The empirical IDR 
$\nu^{-1} \bar F^{-}(0,0)\approx -0.44 \; \mu^{-2}$, indicating that
if $\nu$ and $t$ are kept fixed, the PCAC extrapolation from the chiral
symmetric Adler-Weisberger limit to the chirally broken real world is
minimal. Perhaps not as small as the 2\% single soft pion
Goldberger-Treiman result of section 3.4 nor the ($\approx 5$\%) single
soft pion Adler consistency condition result of section 3.5, but still the PCAC
hypothesis appears to work well in this more stringent test.  Since the
$\pi$N amplitude can be written with the choice of variables $\nu,t;
q^2, q'^2$ or $\nu, q\cdot q';q^2, q'^2$ or indeed some other
combination, the magnitude of the PCAC corrections will depend on which
pair $(\nu, t)$ or $(\nu, \nu_B= q\cdot q'/2m )$ is held fixed during
the extrapolation from $q\rightarrow 0$ to $q^2 =
\mu^2$~\cite{PCACfootnote}. From our experience with (\ref{eq:adlerlet})
in Section 3.5, we argue that holding fixed $(\nu, t)$ (see Fig. 10) is
the correct way to apply PCAC.

Given these empirical values of the on-shell amplitude $\bar F^{+}(\nu
,t)$ and $\nu^{-1} \bar F^{-}(\nu ,t)$, one can now visualize in Fig.
10 the proposed tests of the (isospin-even) PCAC low energy
theorems(LET) labeled Adler LET (\ref{eq:adlerlet})) and Weinberg LET,
the latter the on-shell analogue of  (\ref{eq:Wpt}).  Fig. 10 depicts
the projection onto the hyperspace $\nu = 0$ of the coordinates 
$(\nu,t,q^2,q'^2)$ needed to describe a fully (pion) off-shell
amplitude.  The extrapolations shown are from the soft pion points A and
A' where $\bar F^{+}(q\rightarrow 0) = \bar F^{+}(q'\rightarrow 0) = 0$
and from the double soft Weinberg point 
$\bar F^{+}(q\rightarrow 0,q'\rightarrow 0)= \bar F^{+}(0,0;0,0) =
-\frac{\sigma_N(0)}{f_\pi
^2}$ to the on-pion-mass-shell line, holding fixed $\nu$ and $t$.  The
scale of the various points on the figure and of the isospin-even 
tests is given by the Cheng-Dashen LET  (\ref{eq:cd}) and the (expected
by PCAC) ``anti-Cheng-Dashen" value at the Weinberg point  (\ref{eq:Wpt}). 
  
The IDR amplitude takes the following values on
 the on-mass-shell line:
\begin{equation}
\begin{array}{cccccc}
 \bar{F}^+(0,t=2\mu^2)  &=& \frac{\sigma_N(2\mu^2)}{f_\pi ^2} &+&
	{\cal O}(\mu^4) &\approx +1.35\;\mu^{-1}  \\
\bar{F}^+ (0,t=\ \mu^2)  &=& 0 &+& {\cal O}(\mu^2)  &\approx -0.08\;\mu^{-1} \\
  \bar{F}^+ (0,t=\ \ 0) &=& - \frac{\sigma_N(0)}{f_\pi ^2} &+& {\cal O}(\mu^2)
   &\approx -1.34\;\mu^{-1} 
\label{eq:PCACcorr}
\end{array}
\end{equation}
Let us make a careful distinction between the expected corrections
indicated in (\ref{eq:PCACcorr}).  The ${\cal O}(\mu^4)$ corrections of
the top line are the result of a rigorous on-shell derivation of a Ward
identity~\cite{BPP,CD}.  The putative ${\cal O}(\mu^2)$ corrections of
the lower two lines are what one might expect from the already
discussed corrections to the  Goldberger-Treiman relation, the Adler
zero, and the Adler-Weisberger relation as one goes from the chiral
symmetric world to the world of $\pi$N scattering.  It is those latter
presumed PCAC corrections which we are trying to test.  If one extends
the observed pattern of small PCAC corrections to the bottom line of
(\ref{eq:PCACcorr}), one can interpret $\bar{F}^+ (0,t=0)\approx  -
\bar{F}^+(0,t=2\mu^2)$  as indicating that the $t$ dependence of
$\sigma(t)$ (\ref{eq:sigterm}),  as determined from the $\pi N$
scattering data, is quite small indeed.  The alternative picture of 
$\frac{\sigma_N(2\mu^2)}{f_\pi ^2} -  \frac{\sigma_N(0)}{f_\pi ^2}
\approx 0.25 \;\mu^{-1}$ \cite{Sainio} would demand a quite large PCAC
correction along the lower plane of  Fig. 10 to get back to the
empirical amplitude;  much larger than any other PCAC correction
evaluated in the $\pi$N system or elsewhere (Ref.~\cite{Scadronrev}).
Furthermore, we will see in Fig. 11 that the amplitude $\bar{F}^+(0,t)$
is nearly linear in $t$ in the interval between $t=0$ and $t=2\mu^2$
(with increasing curvature at $t$ approaches the $\pi \pi \rightarrow N
\bar{N}$ pseudo-threshold at $4\mu^2$).   We will return to this issue
after investigating the uniqueness of the IDR amplitudes from which
these conclusions are drawn.

 \setcounter{figure}{9}
\begin{figure}[htpb]
\unitlength1.cm
\begin{picture}(6,12)(+1,4.5)
\includegraphics{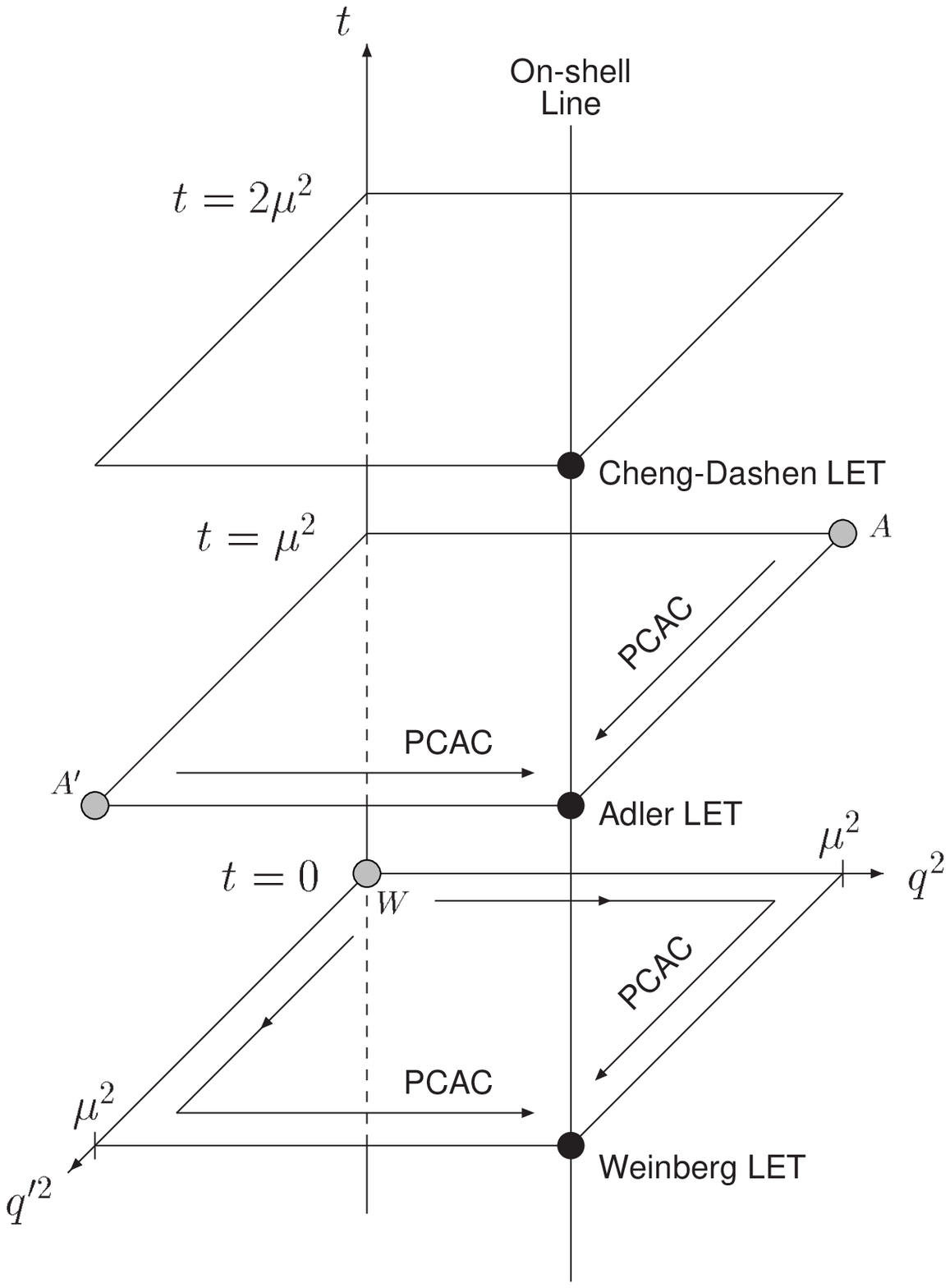}
\end{picture}
\caption{The geometry of the off-mass-shell $\pi$N amplitude 
$\bar{F}^+(\nu,t,q^2,q'^2)$ for $\nu = 0$.}

\end{figure}

  \setcounter{figure}{10}
\begin{figure}[htpb]
\unitlength1.cm
\begin{picture}(5,9)(-15,2.3)
\includegraphics{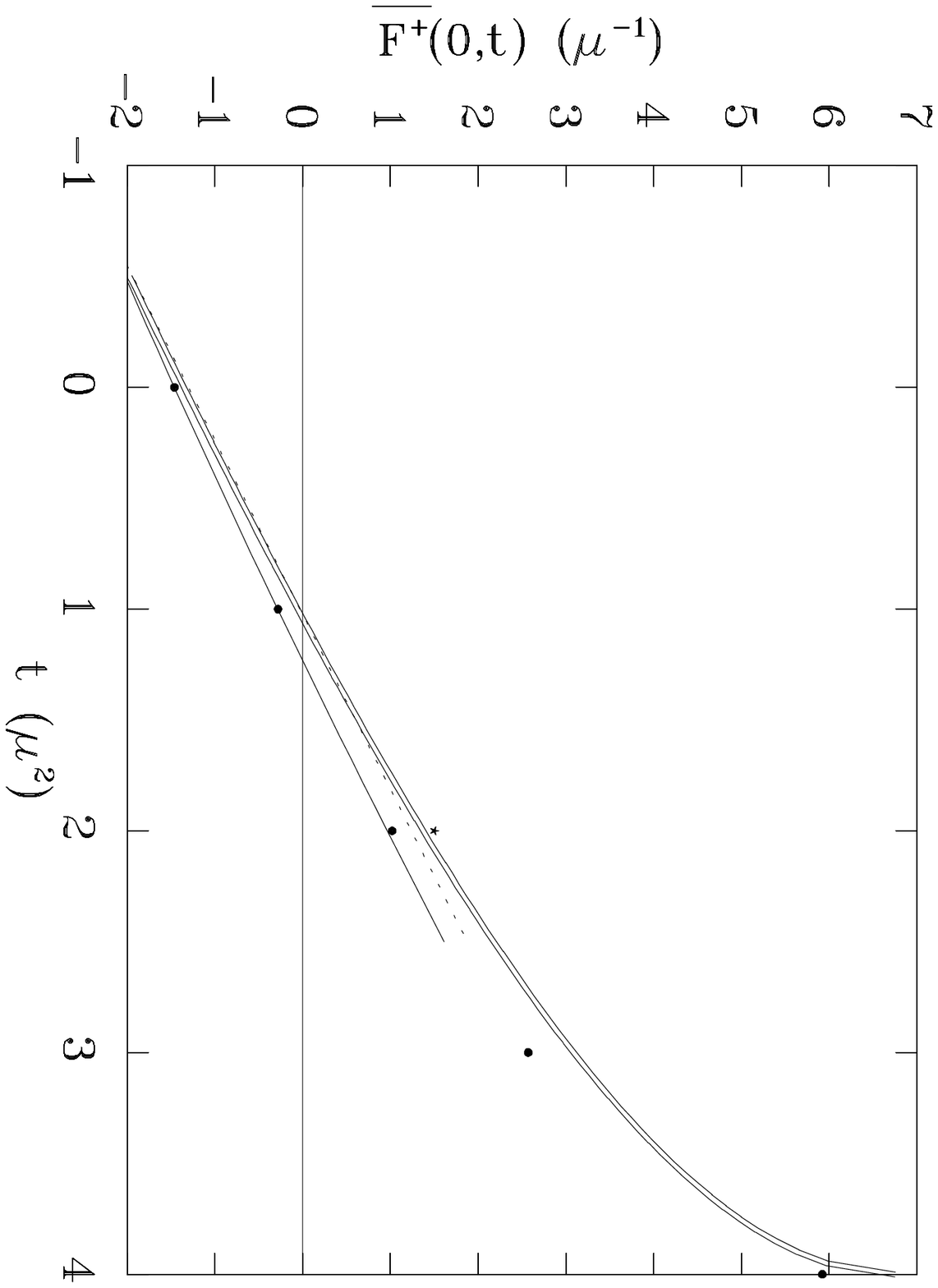}
\end{picture}
\caption{The values of $\bar{F}^+(0,t)$ at the on-mass-shell line of
Fig.10.  The solid line and the filled circles are from the pre-meson
factory data analysed in Ref.~\cite{hohlerbook}.  The double line
corresponds to the IDR analyses of meson factory phaseshifts SM98
(Ref.~\cite{Kaufmann}) and the short dashed line is constructed with the
subthreshold coefficients determined from SM98 with forward dispersion
relations (Ref.~\cite{Pavan}).  The star includes the ``curvature
corrections" to estimate  the latter
amplitude at the Cheng-Dashen point.}
\end{figure}

 The value of the IDR amplitude   $\bar F^{+}(\nu=0, t = 2\mu^2) \approx
 1.35\;\mu^{-1}$ at the Cheng-Dashen point leads to a value of the sigma
 term from (\ref{eq:cd}) of
\[  \sigma_N(2\mu^2) = \bar{F}^+(0,2\mu^2)  {f_\pi ^2}\approx 83 \;\;{\rm
MeV}  \]
 This value is at the high end of a range of 40 MeV to 80 MeV presented
 in 1997 at the MENU97 conference and reviewed there by
 Wagner~\cite{sigmarange}.   An independent application of forward
 dispersion relations to the SM98 partial wave analysis (itself heavily
 influenced  by various sets of dispersion relation
 constraints~\cite{Pavan}) gives  
  \[  \sigma_N(2\mu^2) = \bar{F}^+(0,2\mu^2)  {f_\pi ^2} \approx ( f^+_1
+ 2\mu^2 f^+_2){f_\pi ^2} \approx 77\;\;{\rm MeV}  \]  					
where we use the notation of Appendix A of Ref.~\cite{TM79} for the
H\"{o}hler subthreshold coefficients.  To this the authors of
Ref.~\cite{Pavan} estimate in
various ways a ``curvature correction" to arrive at a final value of 82
to 92 MeV for $\sigma_N(2\mu^2)$.  They  also obtain at the Weinberg LET
$\bar{F}^+ (0,t=0)\equiv f^+_1 = -1.30 \;\mu^{-1}$ which agrees well
with the IDR value.  We have already noted in section 3.5 that the
 Adler zero is closely emulated both by the IDR value 
 $\bar{F}^+ (0,t=\mu^2) \approx -0.08\;\mu^{-1}$ and by the forward
 dispersion relation value $\bar{F}^+ (0,t=\mu^2) = f^+_1 + \mu^2
 f^+_2 = (-1.30 + 1.27) \;\mu^{-1} = -0.03\;\mu^{-1}$.  These two
 different dispersion relation analyses of the same set of phase shifts
 agree well with each other on the subthreshold (but on-shell) line 
 ($\nu = 0, 0\leq t \leq 2\mu^2$) important for tests of PCAC (See
 Fig. 11).  Both analyses include the statements that
 if the authors replace the SM98 phase shifts by the old Karlsruhe phase
 shifts (from the pre-meson factory data) they reproduce the Karlsruhe
 dispersion relation results.  It would seem that this value of the
 sigma term follows from the SM98 phase shifts and is not an artifact of
 a particular type of dispersion theory analysis.  Be warned, however,
 that extrapolations to the
 unphysical Cheng-Dashen point with potential model amplitudes fit to
 (perhaps) different data sets give
 sigma terms at the low end of the MENU97 range and the reader is
 encouraged to continue monitoring the situation, especially the
 CNI-experiment with CHAOS at TRIUMF~\cite{atom}.

 We have established the scale  set by the size of the sigma term, the
 near linearity and 
 change of sign of the empirical amplitude $\bar{F}^+ (0,t)$ in the
 range $0\leq t \leq 2\mu^2$, and the validity of the PCAC hypothesis
 for the off-shell amplitudes $\bar{F}^+ (0,t;q^2,q'^2)$ and 
 $\bar{F}^- (0,t;q^2,q'^2)$.  It remains to discuss the
 $t$ dependence of the first terms in the on-shell current algebra Ward
 identities of Eqs. (\ref{eq:fplus}), (\ref{eq:fminus}), 
   and (\ref{eq:bminus}).  The current algebra terms (49) are simply
   given by the measured electromagnetic form factors.  Given that the
   intrinsic $t$ dependence of $\sigma_N(t)$ is quite small, one can
    set
\begin{equation}
 	\bar{F}^+(\nu,t) = \left (\frac{\sigma_N(2\mu^2)}{f_\pi
	^2}\right )
	[1 + \beta(\frac{t}{\mu^2}-2)] +  
                       C^+(\nu,t),  \label{eq:pnamp}
\end{equation}
where the background amplitude $C^+(\nu,t)$ of (56) is modeled by the
overwhelmingly dominant $\Delta(1232)$ isobar.  The two approaches to
this background amplitude have used dispersion theory for the (over 20)
invariant amplitudes of the axial-vector nucleon amplitude $M^{ij}_{\mu
\nu}$~\cite{ST} or a $\Delta$-propagator field theory model~\cite{OO}.
Both models of the background amplitude give quite similar
results for  $C^+(0,t)$ in the low $t$ regime~\cite{olddata}.  The
dispersion theoretic 
$C^+(\nu,t;q^2 q'^2) \approx c_1\nu^2 + c_2q\cdot q' + {\cal O}(q^4) $
contains an unknown subtraction constant in the $g_{\mu \nu}$ term of
the axial-nucleon amplitude $\bar{M}^+_{\mu \nu}$, which is moved into
the unknown $\beta$ of (ref{eq:pnamp}) and ultimately determined by the
data.

 This particular $t$ dependence of the multiplier  of $\sigma_N$ is
suggested by a low energy expansion similar to the Weinberg amplitude
for low-energy $\pi \pi$ scattering.  This amplitude in the linear
approximation satisfying all current  algebra/PCAC and quark  model
($(\bar{3},3) + (3,\bar{3})$) constraints is \cite{Weinberg66}
\begin{equation}
    T_{\pi \pi} = \frac{1}{f^2_{\pi}}[(s-\mu^2)\delta^{ab}\delta^{cd}
    + (t-\mu^2)\delta^{ac}\delta^{bd} + (u-\mu^2)\delta^{ad}\delta^{bc}]
    + {\cal O}(\mu^{4}),           \label{eq:pipi}
\end{equation}
along with the quark model $\pi \pi$ $\sigma$ term $\sigma_{\pi \pi}= 
\mu^2$.  Generalizations of (\ref{eq:pipi}) to $SU(3)$ pseudoscalar 
meson-meson scattering were worked out by Osborn \cite{osborn} and by Li 
and Pagels \cite{lp}.  In particular, for $\pi P\rightarrow \pi P$ 
scattering, the 
off-shell low-energy generalization of (\ref{eq:pipi}) in the linear 
approximation includes a 
($t$-channel) isospin-even part
\begin{equation}
   T_{PP}^{t-even} = \frac{\sigma_{PP}}{f^2_{\pi}}
	[ (1-\beta_P) (\frac{q^2 + q'^2}{\mu^2} - 1) + 
\beta_P(\frac{t}{\mu^2}-1)]  + {\cal O}(\mu^{4})   \label{eq:psps}    
\end{equation}
for
\begin{equation}
   \beta_{\pi}, \beta_K, \beta_{\eta_8} = 1,\frac{1}{2}, 0 
\;\;\;\;\;{\rm and}\;\;\sigma_{\pi \pi},\sigma_{KK},\sigma_{\eta_8 
\eta_8} = \left (1, \frac{1}{2},\frac{1}{3}\right )\mu^2.   \label{eq:betas}
\end{equation}
In fact the $t$ dependent structure of (\ref{eq:pnamp}) follows from 
(\ref{eq:psps}) (with constant sigma terms) for 
scattering of on-shell pions from a meson target.  Moreover, this
linear (in t) structure of both (\ref{eq:pnamp}) and (\ref{eq:psps}) 
manifests the Adler and Weinberg soft pion theorems.
 
  For $\pi N\rightarrow\pi N$ scattering, however, the fact that the 
nucleon four-momentum cannot become soft means that $\beta_P$ in 
(\ref{eq:pnamp}) cannot be {\it a priori} predicted as it is in 
(\ref{eq:betas}) for meson targets.  Instead
 $\beta$ in (\ref{eq:pnamp}) for on-mass-shell 
$\pi N\rightarrow\pi N$ scattering is fitted to the the IDR curve in
Fig. 11 to find $\beta\approx 0.45$, quite near to 
$\beta_K = 1/2$ in (\ref{eq:betas}),  perhaps reflecting the same 
isospin structure of K and N.  The above  current algebra/PCAC 
analysis in (\ref{eq:pnamp}) and in (\ref{eq:pipi}-\ref{eq:betas}) does 
{\em not} mean that the $\sigma$ term occurring in four-point function $\pi N$ 
scattering has an intrinsic $t$ dependence.  Rather, the linear 
$t$ dependent factor in 
 (\ref{eq:pnamp}) is a PCAC realization of the unknown subtraction constant 
$\beta q'\cdot q \sigma_N $\ , with the $\beta$ determined by the 
Adler and Weinberg LETs.

An alternative ansatz for the $t$ dependence of the sigma term  stems
from the 
 SU(2) linear
$\sigma$ model (L$\sigma$M) with N, $\pi,\sigma$ as elementary
fields~\cite{GML60}.  (For a recent review of the resurgence of interest
in the $\sigma$ meson, see the references in~\cite{sigmaref}.)
Using a pseudoscalar rather than pseudovector $\pi$NN coupling means 
that the t-channel $\sigma$ pole has a background $\bar F^+$amplitude 
 \cite{Schnitzer} proportional to 
$(m^2_\sigma - \mu^2)(m^2_\sigma - t)^{-1} - 1$.  This structure 
automatically complies with the 3 low-energy theorems ; e.\ g.\ it 
vanishes at t=$\mu^2$ as does the Adler zero. 
Thus the  isospin-even 
background $\pi N$ amplitude can be expressed in the L$\sigma$M at $\nu = 
0$ as \cite{Schnitzer}
\begin{equation}
\bar F^+_{L\sigma M} (0,t) = {g^2_{\pi NN}\over m} 
\left[{m^2_\sigma - \mu^2\over m^2_\sigma - t} - 1 \right]
\label{eq:lsm}  
\end{equation}
To obtain a quantitative fit to the empirical amplitude this must be
supplemented by the $\Delta$ contribution. In the dispersion relation
model the amplitude becomes
\begin{equation}
\bar F^+(0,t)= {g^2_{\pi NN}\over m} 
\left[{m^2_\sigma - \mu^2\over m^2_\sigma - t} - 1 \right] + \beta' q
\cdot q' + C^+(\nu,t)\;  \label{eq:fsigma}
\end{equation}
where again the  parameter $\beta'$ shifts  the unknown
subtraction constant in the $\Delta$ contribution to the PCAC
realization  $\beta' q'\cdot q
\sigma_N $\ . A fit which is within the two lines of the double line 
of Fig. 11 for $0 \leq t \leq 2\mu^2$ can be made with $\beta'= 1.44
\mu^{-3}$ and $m_\sigma = 4.68 \mu \approx 653$ MeV~\cite{Beltran}, the
latter a quite reasonable value when compared with the current $\sigma$
meson phenomenology~\cite{sigmaref}.

With these determinations of the $t$ dependence of the sigma term in
(57), we can finally  finish the tests suggested in the last sentence of
Section 3.  That is, how well do the current algebra models describe the
$\pi$N data extrapolated to the subthreshold crescent of Fig. 2?  Away
from the $\nu = 0$ line in the amplitude $\bar{F}^+$, these tests are
given by expanding the isobar modeled $C^{\pm}(\nu,t)$ and $D^{\pm}(\nu,t)$
and comparing the theoretical expansion coefficients  with the empirical
H\"{o}hler coefficients.  This is an old story and the general trends
are summarized in Appendix A of Ref.~\cite{TM79}.  The 28 subthreshold
coefficients are matched very well indeed by the current algebra
amplitudes (57-60), once the  $t$ dependence of the sigma term is fixed
empirically (as above).  The $t$ dependence of the two current algebra
terms in (58) and (60) is indeed given quite well by the isovector
vector current of (13).  The preliminary determinations of the
subthreshold H\"{o}hler coefficients from the meson factory
data~\cite{Kaufmann2} does
not change the qualitative picture given in~\cite{TM79,ST}.  Only the
scale of $\bar F^+(\nu,t)$, set by the size of the sigma term, has
changed with the advent of increasingly more precise $\pi$N data.

The $t$ dependence of the sigma term was suggested by PCAC off-shell
constraints (\ref{eq:psps}) or by the linear sigma model
(\ref{eq:lsm}). The success of the on-shell tests and of the examples of
the PCAC hypothesis suggests a reliable PCAC off-pion-mass shell
extrapolation for the not-so far extrapolations of the $\pi$N amplitude
discussed in the Introduction. However, the shorter range parts of the
two-pion exchange three-body force will be quite different for the $t$
dependence of the sigma term from (\ref{eq:psps}) or (\ref{eq:lsm}). 
The shorter range parts which follow from (\ref{eq:psps}) have been
discussed and compared with those of chiral perturbation theory in
Ref.~\cite{FHvK} which used the techniques of Ref.~\cite{TW95}.  The
implications of (\ref{eq:lsm}) for three-body forces, threshold pion
production, and pion condensation remain to be worked out.\\

I would like to thank the organizers of the Praha Indian-Summer School for
inviting me to this beautiful city.  I am indebted to Michael D. Scadron
for teaching me about current algebra and PCAC and for reading the early
parts of
this manuscript.  I thank William B. Kaufmann for many discussions of
the IDR analysis of the meson factory $\pi$N data.


\begin{thebibliography}{99}
\bibitem{footnote} It is worth noting that Weinberg uses the term ``soft
pions" to mean low energy pions (energies less than about 1200
MeV~\cite{compare}) rather than the technical meaning $q \rightarrow 0$
of the current algebra and PCAC program.  Similarly he uses the term
``current algebra" to mean current algebra plus the smoothness arguments
of PCAC.  Most practitioners of ChPT follow this lead;  hence one reads
statements like ``ChPT subsumes current algebra".  These statements should
be interpreted with Weinberg's nomenclature just described.
\bibitem{phasefootnote} This  convention of Ref. \cite{Scadronrev} chooses a
  non-unique phase factor.  There are other assignments in the
  literature; for example the ``Condon-Shortley" convention is
  $\pi^+ = -(\pi^1 + i \pi^2)/\sqrt{2}$ and  
   $\pi^- = +(\pi^1 - i \pi^2)/\sqrt{2}$.
\end{thebibliography}
\end{document}